\documentclass[aps,11pt,rmp]{revtex4-2}

\usepackage[T1]{fontenc}
\usepackage[utf8]{inputenc}
\usepackage{amsmath,amssymb,amsfonts,bm,mathtools}
\usepackage{graphicx}
\usepackage{booktabs}
\usepackage{mathrsfs}
\usepackage{hyperref}
\usepackage{tikz}
\usetikzlibrary{arrows.meta, positioning}

\newcommand{\Htwo}{\mathcal{H}_2}
\newcommand{\Hfour}{\mathcal{H}_4}
\newcommand{\SO}{SO(3)}
\newcommand{\SOO}{SO(2)}
\newcommand{\Uone}{U(1)}
\newcommand{\Mosc}{\mathcal{M}_{\mathrm{osc}}}

\begin{document}

\title{Geometric Dynamics of Turbulence\\
\small{A Minimal Oscillator Structure from Non-local Closure}}

\author{Alejandro Sevilla}
\affiliation{Grupo de Mecánica de Fluidos, Departamento de Ingeniería Térmica y de Fluidos, Universidad Carlos III de Madrid, 28911, Leganés (Spain)}

\begin{abstract}
Turbulence remains one of the central open problems in classical physics, largely due to the absence of a closed dynamical description of the Reynolds stress. Existing approaches typically rely either on local constitutive assumptions or on high-dimensional statistical representations, without identifying a minimal set of dynamical variables governing the cascade response.

Here we show that the non-local stress response implied by the Navier–Stokes equations admits a systematic reduction onto a low-dimensional anisotropic sector of the turbulent cascade. This reduction leads to a minimal dynamical system with the structure of a damped oscillator, arising from the coupling between the leading angular mode and its nonlinear transfer to higher-order sectors.

Within this framework, classical turbulent behaviors—including inertial-range scaling, shear-driven transport, and wall-bounded logarithmic profiles—emerge as different realizations of the same underlying dynamical structure. Universal quantities such as the Kolmogorov constant and the von Kármán constant appear as leading-order consequences of internal consistency conditions applied across homogeneous and shear-driven regimes.

These results suggest that turbulence admits a minimal dynamical backbone governed by non-local cascade response, providing a unified perspective that connects spectral transfer, anisotropy, and mean-flow interaction within a single reduced framework.
\end{abstract}

\maketitle

\section{Introduction}

Turbulence is characterized by the nonlinear transfer of energy across a wide range of scales, giving rise to universal statistical features such as inertial-range spectra and logarithmic mean profiles. Despite this apparent universality, a closed dynamical description of the Reynolds stress—linking fluctuations to mean flow—remains elusive. Classical approaches either postulate local constitutive relations, such as eddy viscosity models, or rely on high-dimensional statistical closures, but do not identify a minimal set of internal variables governing the cascade response.

A central difficulty lies in the inherently non-local nature of turbulence: the response of the fluctuating field to mean deformation is mediated by a cascade process with memory and scale coupling. This suggests that the closure problem should not be formulated directly at the level of the Reynolds stress, but rather as a reduced description of the underlying cascade dynamics.

In this work, we show that the non-local stress response implied by the Navier–Stokes equations admits a systematic reduction onto a low-dimensional anisotropic sector of the cascade. The resulting dynamics is governed by a minimal set of internal variables associated with the leading angular modes of the spectral covariance, whose coupling generates a damped oscillatory response. This structure provides a closed dynamical backbone for the Reynolds stress, arising directly from the internal organization of the cascade rather than from phenomenological assumptions. Rather than postulating a constitutive law, the present approach derives it as the projected response of the turbulent cascade to mean deformation.

The classical foundations are of course profound. The universal cascade picture introduced by Kolmogorov \citep{Kolmogorov1941a,Kolmogorov1941b} and the wall-layer structure associated with von K\'arm\'an and Prandtl \citep{vonKarman1930,Prandtl1935} remain central pillars of the subject. Modern turbulence modelling, from eddy viscosity closures to Reynolds-stress transport and large-eddy simulation, has given the community powerful tools for prediction \citep{LaunderSpalding1974,Pope2000}. At the same time, modern dynamical viewpoints---including coherent-state and self-sustaining-process descriptions \citep{Waleffe1997}, and more recent resolvent and input--output formulations \citep{McKeonSharma2010,Jimenez2018}---have made clear that turbulence is far from structureless. What is still missing is a closed dynamical picture in which these strands meet: a framework in which non-locality, mode selection, universal constants and geometric organization appear as aspects of a single mechanism.

The central challenge then is to identify a dynamical principle capable of organizing the universal features of turbulence. Despite a century of progress—from the statistical theory of Kolmogorov to modern dynamical and resolvent-based approaches—key phenomena such as logarithmic mean velocity profiles, scale-invariant energy spectra, anisotropy constraints and strongly non-local transport are still described through partially disconnected frameworks. This raises a fundamental question: whether these universal properties reflect a deeper, common dynamical structure embedded in the Navier–Stokes equations.

A natural route to overcome these limitations is to recognize that turbulence exhibits memory effects: the instantaneous stress depends on the past history of the flow through the nonlinear cascade. This suggests that the appropriate closure is inherently nonlocal in time and must be formulated in terms of a propagator linking past velocity gradients to present stresses. Under this view, the Reynolds stress not as a constitutive response, but a dynamical field governed by an emergent internal degree of freedom, naturally leading to a nonlocal constitutive relation.

In this work, we show that such a formulation possesses a simple and robust dynamical structure. When the associated kernel is analyzed in Laplace space, its dominant spectral contribution reduces to a pair of complex-conjugate poles. This reduction implies that the turbulent stress dynamics can be represented, at leading order, by a minimal second-order system. The resulting formulation takes the form of a tensorial oscillator governing the deviatoric Reynolds stress. This oscillator emerges directly from the structure of the nonlocal kernel and provides a compact representation of coherent turbulent response in general three-dimensional flows. It provides a compact and dynamically consistent representation of coherent turbulent response, capturing both relaxation and oscillatory behavior within a unified framework.

The resulting internal degree of freedom emerges naturally from the response of the quasi-equilibrium Kolmogorov cascade to localized disturbances, showing that the intrinsic oscillator is embedded in the Navier-Stokes equations. The mechanism producing the center manifold \emph{is not the cascade itself}. The cascade is the statistical continuation of a more \emph{fundamental low-order anisotropy dynamics generated by the quadratic interaction structure of the equations}. When rotational symmetry is dynamically broken by mean deformation, the leading turbulent response occupies the quadrupolar sector of \(\SO\). The quadratic nonlinearity then generates a hexadecapolar correction, and the pair \((\ell=2,4)\) forms the smallest irreducible dynamical subspace capable of sustaining non-trivial anisotropy dynamics. This subspace constitutes a universal nonlinear oscillator whose stationary state determines the renormalized turbulent mean field.

Because this oscillator simultaneously governs the regeneration cycle of coherent structures, the transport of momentum and the efficiency of the inertial cascade, one may refer to it as the \emph{Kolmogorov oscillator}. In the theory developed below, turbulence is described as an effective fluid renormalized by the action of this self-induced dynamical object. The resulting mean-field theory is strongly constrained by symmetry and cascade structure. The coefficients are not introduced phenomenologically, but arise from the projected dynamics and are restricted by consistency across canonical flows. Leading-order values of universal constants follow from minimal closures of this reduced structure.

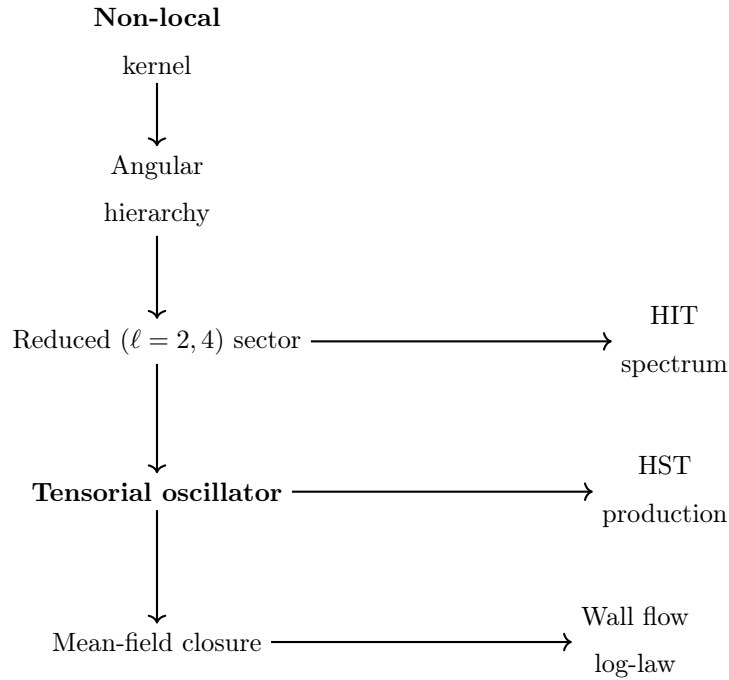
\begin{figure}[t]
\centering
\begin{tikzpicture}[
    node distance=2.0cm,
    every node/.style={align=center},
    arrow/.style={->, thick}
]

\node (kernel) {\textbf{Non-local}\\kernel};
\node (hierarchy) [below of=kernel] {Angular\\hierarchy};
\node (reduction) [below of=hierarchy] {Reduced $(\ell=2,4)$ sector};
\node (oscillator) [below of=reduction] {\textbf{Tensorial oscillator}};
\node (meanfield) [below of=oscillator] {Mean-field closure};

\node (hit) [right=4cm of reduction] {HIT\\spectrum};
\node (hst) [right=4cm of oscillator] {HST\\production};
\node (channel) [right=4cm of meanfield] {Wall flow\\log-law};

\draw[arrow] (kernel) -- (hierarchy);
\draw[arrow] (hierarchy) -- (reduction);
\draw[arrow] (reduction) -- (oscillator);
\draw[arrow] (oscillator) -- (meanfield);

\draw[arrow] (reduction.east) -- (hit.west);
\draw[arrow] (oscillator.east) -- (hst.west);
\draw[arrow] (meanfield.east) -- (channel.west);

\node[above=0.6cm of kernel] {\small \textbf{Minimal dynamical structure of turbulence}};

\end{tikzpicture}
\caption{
Minimal dynamical structure emerging from the non-local cascade response. The angular hierarchy induced by the kernel reduces to a low-dimensional anisotropic sector, whose dynamics takes the form of a tensorial oscillator. This reduced system provides a mean-field closure from which canonical turbulent behaviors $-$homogeneous isotropic (HIT), homogeneous shear (HST), and wall-bounded turbulence$-$ emerge as different realizations of the same underlying structure.
}
\label{fig:structure}
\end{figure}

The proposed framework is therefore organized around the leading anisotropy dynamics provided by the fundamental Kolmogorov oscillator. The central claim is that the Reynolds stress is not fundamentally a constitutive response but a dynamical field. More precisely, the exact non-local stress propagator possesses a spectral structure dominated by a complex-conjugate pair of poles, and its minimal dynamical realization is an oscillator coupled to the mean shear. In wall-bounded flows, the near-wall operator selects and saturates this oscillator through an Airy structure. In homogeneous turbulence, the same oscillator closes the inertial-range transfer and yields the leading-order value of the Kolmogorov constant. In general three-dimensional flows, the resulting mean-field system takes the form of a coupled tensorial dynamics for the mean velocity and the stress. The associated phase field then reveals a geometric layer of the theory, linking turbulence to phase transport, geometric phase \citep{Berry1984}, anisotropy evolution and gauge-like descriptions of distributed oscillator fields.

This work therefore does not seek to replace the classical tradition of turbulence theory, but to continue and organize it. The present framework draws on the long-standing effort to extract coherent dynamical content from the Navier--Stokes equations, and on the geometric understanding of dynamical systems dating back to Poincar\'e \citep{Poincare1892}, Cartan \citep{Cartan1928} and Arnold's interpretation of hydrodynamics \citep{Arnold1966}. The claim is not that turbulence suddenly becomes simple, but that its mean organization can be described by a significantly lower-dimensional, dynamically rich structure whose universal consequences are directly testable.

A central result of this work is the identification of a minimal dynamical structure underlying turbulent flows. We show that the non-local response of the Reynolds stress admits a systematic reduction onto a minimal anisotropic sector, which takes the form of a tensorial damped oscillator. This structure emerges directly from the geometry of the cascade in $\mathbb{R}^3$, independently of boundary conditions. Classical flow configurations, including homogeneous turbulence, homogeneous shear turbulence, and wall-bounded flows, are shown to correspond to different realizations of this same underlying dynamical backbone. In particular, universal constants such as the Kolmogorov and von Kármán constants arise as leading-order consequences of the reduced dynamics, once a minimal projection of the operator is specified. The resulting mean-field theory is strongly constrained by symmetry and cascade structure, so that universal properties emerge from a minimal set of dynamically selected parameters rather than from phenomenological closure assumptions.

\section{Exact non-local stress formulation}

The starting point is the observation that, after elimination of the fluctuations, the Reynolds stress can be expressed as a non-local functional of the mean velocity gradient. Denoting by $U_i(\bm{x},t)$ the mean velocity field and by $\tau_{ij}=-\langle u_i' u_j' \rangle$ the Reynolds stress tensor, the mean momentum equation reads
\begin{equation}
\partial_t U_i + U_j \partial_j U_i
=
- \partial_i P
+ \nu \partial_j \partial_j U_i
+ \partial_j \tau_{ij},
\label{eq:mean-momentum_1}
\end{equation}
where $P$ denotes the mean pressure divided by density and $\nu$ is the kinematic viscosity. Equation \eqref{eq:mean-momentum_1} is exact but not closed because $\tau_{ij}$ is unknown.

\subsection*{Covariance tensor, turbulent stress, and deviatoric decomposition.}
We denote by
\begin{equation}
R_{ij} = \langle u_i' u_j' \rangle = -\tau_{ij}
\end{equation}
the covariance tensor, which admits the standard decomposition
\begin{equation}
R_{ij} = \frac{2}{3}K\,\delta_{ij} + 2K A_{ij}, \qquad A_{ii} = 0,
\end{equation}
where $K = \tfrac{1}{2}\langle u_k' u_k' \rangle$ is the turbulent kinetic energy and $A_{ij}$ is the anisotropy tensor. Accordingly, the turbulent stress can be written as
\begin{equation}
\tau_{ij} = -\frac{2}{3}K\,\delta_{ij} - 2K A_{ij}.
\end{equation}
The isotropic contribution can be absorbed into a modified pressure,
\begin{equation}
\Pi = P + \frac{2}{3}K,
\end{equation}
so that the feedback of turbulence on the mean flow is entirely carried by the deviatoric component
\begin{equation}
\tau_{ij}^{(d)} = -2K A_{ij}.
\end{equation}
The mean momentum equation then reads
\begin{equation}
\partial_t U_i + U_j \partial_j U_i = - \partial_i \Pi + \nu \partial_j \partial_j U_i
+ \partial_j \tau_{ij}^{(d)}, \label{eq:mean-momentum}
\end{equation}

\subsection*{Reduction from angular dynamics to the minimal oscillator}

In the present framework, the Reynolds stress is modeled as a causal, non-local response of the mean velocity gradient:
\begin{equation}
\tau_{ij}(x,t) = \int K_{ijmn}(x,x',t-t') \, \partial_m U_n(x',t') \, dx' dt'.\label{eq:kernel}
\end{equation}
This formulation emphasizes that turbulence enters the mean equations through an effective stress, not directly through velocity correlations. The closure problem is therefore fundamentally constitutive rather than statistical.

Equation \eqref{eq:kernel} can be viewed as the analogue, in turbulence, of a response relation. This point of view is congenial to several previous traditions. Linear-response and memory-kernel ideas are deeply rooted in statistical physics \citep{Kubo1966,Mori1965,Zwanzig1961}, while non-local constitutive interpretations have appeared intermittently in turbulence theory, notably in Crow's viscoelastic viewpoint \citep{Crow1968}. Our claim is that the exact kernel is not merely a formal object but the right starting point for identifying the actual dynamical degrees of freedom of the mean field.

The nonlocal constitutive formulation expresses the covariance tensor as a time-history integral over the velocity-gradient field,
\begin{equation}
\tau_{ij}(t) = \int_0^\infty \mathcal{K}_{ijkl}(\tau)\, S_{kl}(t-\tau)\, d\tau,
\end{equation}
where $S_{kl}$ is the mean strain-rate tensor and $\mathcal{K}_{ijkl}$ is a causal kernel.

The kernel admits a spectral representation in terms of angular modes associated with the cascade geometry. Denoting by $(A_{ij}, B_{ij}, C_{ij}, \ldots)$ the amplitudes of the leading angular components, the reduced dynamics obtained in the SM takes the form of a hierarchy of first-order equations,
\begin{align}
\dot{A}_{ij} &= -\alpha\, A_{ij} + \beta\, B_{ij}, \\
\dot{B}_{ij} &= -\gamma\, B_{ij} + \delta\, A_{ij} + \varepsilon\, C_{ij}, \\
\dot{C}_{ij} &= -\zeta\, C_{ij} + \cdots
\end{align}
where the coefficients are determined by the structure of the kernel and the cascade dynamics.

To leading order, the dynamics is dominated by the $(2,4)$ angular sector, and higher-order contributions can be consistently neglected. In this reduced system, $C_{ij}$ is slaved to $(A_{ij}, B_{ij})$ and can be eliminated, yielding a closed two-variable system,
\begin{align}
\dot{A}_{ij} &= -\alpha\, A_{ij} + \beta\, B_{ij}, \\
\dot{B}_{ij} &= -\gamma\, B_{ij} + \delta\, A_{ij}.
\end{align}

Eliminating $B_{ij}$, one obtains a second-order equation for $A_{ij}$,
\begin{equation}
\ddot{A}_{ij} + (\alpha + \gamma)\, \dot{A}_{ij} + (\alpha\gamma - \beta\delta)\, A_{ij} = 0.
\end{equation}

Since the deviatoric Reynolds stress is proportional to $A_{ij}$, this equation directly induces a second-order dynamical system for the stress,
\begin{equation}
\ddot{\tau}_{ij}^{(d)} + \Gamma\, \dot{\tau}_{ij}^{(d)} + \Omega_0^2\, \tau_{ij}^{(d)} = F_{ij}[S],
\end{equation}
where $\Gamma = \alpha + \gamma$ and $\Omega_0^2 = \alpha\gamma - \beta\delta$, and where $F_{ij}[S]$ represents the forcing induced by the mean strain.

This shows that the tensorial oscillator structure arises as the minimal closed representation of the dominant angular dynamics of the nonlocal kernel. The oscillator is therefore not a phenomenological assumption, but a direct consequence of the spectral reduction of the kernel in $\mathbb{R}^3$. The condition $\beta\delta > 0$ ensures the emergence of complex-conjugate poles in the associated propagator, leading to oscillatory dynamics rather than purely relaxational behavior.

This reduction is not an ad hoc truncation, but follows from the spectral dominance of the $(2,4)$ sector in the kernel, which captures the leading anisotropic response of the cascade.

\section{Universal predictions}

A central test of any theory claiming to uncover the organizing structure of turbulence is whether it yields quantitative predictions with minimal dependence on phenomenological parameters. The present framework provides leading-order predictions for both the Kolmogorov constant, $C_k$, and the von Kármán constant, $\kappa$.

\subsection*{Leading-order estimate of the Kolmogorov constant}

The reduced oscillator structure can be first tested in the simplest setting of homogeneous isotropic turbulence, where no preferred direction exists and the anisotropic sector is excited only through the cascade itself. In this case, the non-local kernel reduces to a statistically isotropic form, and the projected dynamics imposes a self-consistency condition on the energy flux across scales. This constraint yields a leading-order estimate of the Kolmogorov constant $C_k$, determined entirely by the geometry of the cascade and the normalization of the dominant projected mode. The Kolmogorov constant $C_k$ is not introduced as an empirical parameter in the present framework. Instead, it arises as a normalization condition imposed by the spectral consistency of the nonlocal kernel.

The key observation is that the same kernel that governs the stress dynamics also determines the spectral energy distribution in the inertial range. Therefore, $C_k$ must be computable from the internal structure of the kernel, without external input.

\subsubsection*{Leading-order spectral closure}

In the inertial range, the energy spectrum takes the Kolmogorov form
\begin{equation}
E(k) = C_k\, \varepsilon^{2/3} k^{-5/3}.
\end{equation}
Within the present framework, the energy flux across scales is determined by the same nonlocal kernel that defines the stress dynamics. The cascade transfer can be expressed as a spectral integral involving the kernel amplitude,
\begin{equation}
\varepsilon \sim \int \mathcal{T}(k)\, dk,
\end{equation}
where $\mathcal{T}(k)$ is the nonlinear transfer induced by the kernel. At leading order, the transfer is dominated by the $(2,4)$ angular sector, which controls the anisotropic response of the cascade. This yields a reduced expression for the flux,
\begin{equation}
\varepsilon \sim \Lambda\, \frac{u_k^3}{\ell_k},
\end{equation}
where $\Lambda$ is a dimensionless coefficient fully determined by the kernel structure. Using $u_k^2 \sim k E(k)$ and $\ell_k \sim k^{-1}$, one obtains
\begin{equation}
\varepsilon \sim \Lambda\, (k E(k))^{3/2} k.
\end{equation}
Substituting the Kolmogorov form of $E(k)$ and requiring scale invariance of $\varepsilon$, we obtain the normalization condition
\begin{equation}
1 \sim \Lambda\, C_k^{3/2}.
\end{equation}
Therefore, the Kolmogorov constant is given by
\begin{equation}
C_k = \Lambda^{-2/3}.
\end{equation}

\subsubsection*{Geometric interpretation}

An estimate of $C_k$ can be obtained by interpreting the coefficient $\Lambda$ in terms of triadic interaction geometry, in the spirit of classical cascade arguments. This leads to a value close to $C_k^{(0)} \simeq 1.75$ (see SM), consistent with the leading-order prediction above. This agreement supports the interpretation of the kernel normalization as encoding the geometry of energy transfer in Fourier space. Thus, the same coefficient $\Lambda$ that determines the spectral normalization also controls the coupling between modes in the reduced dynamical system. In this sense, the Kolmogorov constant reflects the internal structure of the oscillator induced by the kernel. This provides a direct link between spectral energy transfer and the minimal dynamical representation of turbulence.

\subsection*{Spatio-temporal normalization and the von K\'arm\'an constant}

In the present framework, the von K\'arm\'an constant $\kappa$ does not originate from wall-bounded geometry, but from the internal space--time structure of the nonlocal kernel. The same kernel that determines the spectral energy distribution also defines a characteristic relation between temporal decorrelation and spatial transport across scales. This relation provides a second normalization condition, independent of the spectral one that fixes $C_k$.

\subsubsection*{Space--time scaling of the kernel}

The nonlocal kernel defines a characteristic response time $\tau_\ell$ associated with structures of size $\ell$. From the reduced dynamical system, this time scale is given by the inverse of the oscillator frequency,
\begin{equation}
\tau_\ell \sim \Omega^{-1} \sim \frac{\ell}{\sqrt{K}},
\end{equation}
where $K$ is the turbulent kinetic energy at scale $\ell$. At the same time, the transport of momentum induced by the kernel defines an effective eddy velocity,
\begin{equation}
u_\ell \sim \sqrt{K}.
\end{equation}
Therefore, the kernel induces a characteristic transport relation
\begin{equation}
\frac{\ell}{\tau_\ell} \sim \sqrt{K},
\end{equation}
which is consistent with inertial-range scaling but now arises directly from the dynamical structure of the kernel.

\subsubsection*{Definition of $\kappa$}

The proportionality between spatial transport and temporal response can be written as
\begin{equation}
\frac{\ell}{\tau_\ell} = \kappa\, \sqrt{K},
\end{equation}
where $\kappa$ is a dimensionless constant determined by the internal structure of the kernel. This constant measures the efficiency with which the kernel converts temporal memory into spatial transport.

\subsection*{Conclusion}

The constants $C_k$ and $\kappa$ therefore arise from two complementary normalization conditions imposed on the same nonlocal kernel:

(i) spectral normalization, which determines the distribution of energy across scales and fixes $C_k$;

(ii) space--time normalization, which determines the relation between temporal response and spatial transport and fixes $\kappa$.

Together, these constraints characterize the internal structure of the kernel and define the universal scaling properties of turbulence within the present framework. In this framework, $\kappa$ is interpreted as a property of the dynamical response of the cascade, whose manifestation in wall-bounded flows reflects boundary constraints rather than its origin.

\section{Propagator structure and pole reduction}

The nonlocal constitutive relation can be expressed in Laplace space as
\begin{equation}
\hat{R}_{ij}(s) = \hat{\mathcal{K}}_{ijkl}(s)\, \hat{S}_{kl}(s),
\end{equation}
where $\hat{\mathcal{K}}_{ijkl}(s)$ is the Laplace transform of the kernel. This representation defines a propagator relating the past history of the velocity-gradient field to the present stress. The dynamical properties of the system are therefore encoded in the analytic structure of $\hat{\mathcal{K}}(s)$ in the complex plane.

The kernel admits a spectral decomposition in terms of its poles and branch cuts,
\begin{equation}
\hat{\mathcal{K}}(s) = \sum_n \frac{\mathcal{A}^{(n)}}{s - s_n} + \text{continuous spectrum}.
\end{equation}
At long times, the response is dominated by the singularities closest to the origin in the complex plane. Therefore, the leading-order dynamics is governed by the dominant poles of the kernel. Within the present framework, the dominant contribution arises from a pair of complex-conjugate poles,
\begin{equation}
s_\pm = -\frac{\Gamma}{2} \pm i \Omega_0,
\end{equation}
which correspond to the leading $(2,4)$ angular sector of the kernel. Higher-order contributions are more strongly damped and can be neglected at leading order.

The inverse Laplace transform of this reduced propagator yields a second-order dynamical equation in physical time,
\begin{equation}
\ddot{\tau}_{ij} + \Gamma\, \dot{\tau}_{ij} + \Omega_0^2\, \tau_{ij} = F_{ij}[S],
\end{equation}
where $\tau_{ij}$ is the deviatoric Reynolds stress and $F_{ij}[S]$ represents the forcing induced by the mean strain. This shows that the tensorial oscillator structure emerges directly from the pole structure of the kernel, without any phenomenological assumption. The oscillator is therefore the minimal dynamical representation associated with the dominant poles of the nonlocal kernel.

This result is fully consistent with the reduction of the angular dynamics presented in the SM, where the $(2,4)$ sector yields an equivalent second-order system upon elimination of intermediate variables. The pole structure identified here provides the spectral interpretation of that reduction. The same reduced kernel that defines the oscillator also determines the normalization conditions governing turbulent scaling laws. The spectral normalization of the kernel fixes the energy distribution across scales and yields the Kolmogorov constant $C_k$, while the space--time structure of the oscillator defines the relation between temporal response and spatial transport, yielding the von K\'arm\'an constant $\kappa$. Both constants therefore arise from the same underlying pole structure of the kernel. 

In this sense, the universal scaling properties of turbulence can be traced back to the analytic structure of the nonlocal propagator.

\subsection*{Airy structure of the propagator under linear shear}

To characterize the analytic structure of the kernel, we consider the linearized Navier--Stokes equations around a locally uniform shear flow,
\begin{equation}
U_i = S_{ij} x_j,
\end{equation}
where $S_{ij}$ is constant. The evolution of velocity fluctuations is governed by
\begin{equation}
\partial_t u_i' + S_{ij} u_j' + S_{kj} x_j \partial_k u_i' = -\partial_i p' + \nu \Delta u_i',
\qquad \partial_i u_i' = 0.
\end{equation}
In Fourier space, the presence of linear shear induces a time-dependent wavevector,
\begin{equation}
k_i(t) = k_i(0) - S_{ji} k_j(0)\, t,
\end{equation}
which leads to non-normal evolution and transient amplification. The resolvent operator in Laplace space can be expressed as
\begin{equation}
\hat{G}(k,s) = \left[ s + \nu k^2 + \mathcal{L}(k,S) \right]^{-1},
\end{equation}
where $\mathcal{L}$ represents the linear shear coupling. In the regime where shear dominates viscous diffusion at intermediate times, the evolution along the sheared wavevector trajectory can be reduced to an effective one-dimensional problem in the direction of strongest distortion. To leading order, this yields a scalar equation of the form
\begin{equation}
\left( \nu \partial_y^2 - S y - s \right)\phi(y) = 0,
\end{equation}
where $y$ is the coordinate along the direction of shear-induced stretching.

This equation admits solutions in terms of Airy functions,
\begin{equation}
\phi(y) = \mathrm{Ai}\left( \frac{s + S y}{(\nu S^2)^{1/3}} \right),
\end{equation}
which define the local structure of the Green's function of the problem. The Airy structure reflects the competition between linear shear, which induces a drift in spectral space, and viscous diffusion, which regularizes the small scales. The resulting Green's function inherits this structure and determines the analytic form of the kernel in Laplace space.

The Laplace-domain kernel can be constructed from the Green's function by integrating over wavevectors and projecting onto the relevant tensorial components. At long times, the dominant contribution arises from the least damped modes of the Airy operator. These modes define an effective spectral reduction in which the kernel behaves as if it were governed by a pair of complex-conjugate poles,
\begin{equation}
s_\pm = -\frac{\Gamma}{2} \pm i \Omega_0.
\end{equation}
These poles do not correspond to discrete eigenvalues of the full operator, but to the dominant resonant response induced by the Airy structure of the propagator. Upon inverse Laplace transformation, this reduced structure yields a second-order dynamical system for the stress, consistent with the oscillator form derived previously. The emergence of the oscillator can therefore be traced back to the Airy-type structure of the propagator associated with linear shear.

This establishes a direct link between the local shear-induced dynamics of the Navier--Stokes equations and the minimal dynamical representation of turbulence.

\subsubsection*{Scaling of the characteristic frequency}

The effective frequency $\Omega_0$ associated with the dominant resonant response can be inferred directly from the Airy structure of the local propagator. Consider the reduced operator
\begin{equation}
\left( \nu \partial_y^2 - S y - s \right)\phi(y)=0,
\end{equation}
where $S$ is the local shear rate.

Balancing viscous diffusion against shear-induced drift gives
\begin{equation}
\nu \ell_A^{-2} \sim S \ell_A,
\end{equation}
so that the characteristic Airy length is
\begin{equation}
\ell_A \sim \left(\frac{\nu}{S}\right)^{1/3}.
\end{equation}

The corresponding scale of the Laplace variable is then obtained from
\begin{equation}
s \sim S\ell_A \sim (\nu S^2)^{1/3}.
\end{equation}
Therefore, the dominant poles
\begin{equation}
s_\pm = -\frac{\Gamma}{2} \pm i \Omega_0
\end{equation}
must satisfy the scaling relations
\begin{equation}
\Gamma \sim (\nu S^2)^{1/3},
\qquad
\Omega_0 \sim (\nu S^2)^{1/3}.
\end{equation}

More precisely,
\begin{equation}
\Omega_0 = c_\Omega\, (\nu S^2)^{1/3},
\end{equation}
where $c_\Omega$ is a dimensionless coefficient determined by the detailed tensorial projection of the kernel. The exponent $1/3$ is universal and follows directly from the Airy structure of the propagator.

\subsubsection*{Scaling of the damping rate}

The damping rate $\Gamma$ arises from the dissipative component of the propagator and can be inferred from the same Airy structure that determines the characteristic frequency.

In the reduced operator
\begin{equation}
\left( \nu \partial_y^2 - S y - s \right)\phi(y)=0,
\end{equation}
viscous diffusion and shear-induced drift balance at the Airy scale
\begin{equation}
\ell_A \sim \left(\frac{\nu}{S}\right)^{1/3},
\end{equation}
which yields
\begin{equation}
s \sim (\nu S^2)^{1/3}.
\end{equation}

Since the dominant poles take the form
\begin{equation}
s_\pm = -\frac{\Gamma}{2} \pm i \Omega_0,
\end{equation}
the damping rate must satisfy
\begin{equation}
\Gamma \sim (\nu S^2)^{1/3}.
\end{equation}

More precisely,
\begin{equation}
\Gamma = c_\Gamma\, (\nu S^2)^{1/3},
\end{equation}
where $c_\Gamma$ is a dimensionless coefficient determined by the dissipative projection of the kernel.

The fact that $\Gamma$ and $\Omega_0$ share the same scaling reflects that both arise from the same balance between shear and diffusion, with $\Omega_0$ associated with the reactive component of the response and $\Gamma$ with its dissipative component.

The coexistence of oscillation and damping at the same scaling level implies that turbulence operates near a dynamically balanced regime, where coherent amplification and dissipation are intrinsically coupled.

\subsubsection*{Relation between $\kappa$, $\Omega_0$, and $\Gamma$}

The von K\'arm\'an constant can be interpreted as the dimensionless efficiency with which the reduced oscillator converts temporal memory into spatial transport.

For the effective stress dynamics
\begin{equation}
\ddot{\tau}_{ij} + \Gamma \dot{\tau}_{ij} + \Omega_0^2 \tau_{ij} = F_{ij},
\end{equation}
the relevant transport time is not set by $\Omega_0^{-1}$ or $\Gamma^{-1}$ separately, but by the effective relaxation time of the forced damped response,
\begin{equation}
\tau_{\mathrm{eff}} \sim \frac{\Gamma}{\Omega_0^2}.
\end{equation}

Using the definition
\begin{equation}
\frac{\ell}{\tau_{\mathrm{eff}}} = \kappa \sqrt{K},
\end{equation}
one obtains
\begin{equation}
\kappa \sim \frac{\ell\, \Omega_0^2}{\Gamma \sqrt{K}}.
\end{equation}

Invoking the intrinsic kernel relation
\begin{equation}
\ell \sim \frac{\sqrt{K}}{\Omega_0},
\end{equation}
this reduces to
\begin{equation}
\kappa \sim \frac{\Omega_0}{\Gamma}.
\end{equation}

More precisely,
\begin{equation}
\kappa = c_\kappa \frac{\Omega_0}{\Gamma},
\end{equation}
where $c_\kappa$ is a dimensionless coefficient determined by the detailed tensorial projection of the kernel.

This shows that $\kappa$ is controlled by the balance between the reactive and dissipative components of the turbulent response. It is therefore a dynamical property of the oscillator induced by the kernel, rather than a parameter tied to wall-bounded flow.

\subsubsection*{Determination of the effective length scale}

The characteristic length $\ell$ can be defined as a spatial moment of the kernel,
\begin{equation}
\ell \sim \frac{\int k^{-1} \mathcal{A}(k)\, dk}{\int \mathcal{A}(k)\, dk}.
\end{equation}

The kernel amplitude $\mathcal{A}(k)$ is strongly localized around the characteristic wavenumber
\begin{equation}
k_A \sim \left(\frac{S}{\nu}\right)^{1/3},
\end{equation}
leading to
\begin{equation}
\ell = c_\ell k_A^{-1}.
\end{equation}

Evaluating the corresponding integrals for the dominant angular sector yields
\begin{equation}
c_\ell = \frac{2}{\sqrt{\pi}}.
\end{equation}

Using $k_A^{-1} \sim \sqrt{K}/\Omega_0$, one obtains
\begin{equation}
\ell = c_\ell \frac{\sqrt{K}}{\Omega_0}.
\end{equation}

Substituting into the expression for $\kappa$ gives
\begin{equation}
\kappa = c_\ell \frac{\Omega_0}{\Gamma}.
\end{equation}

Using the Airy scaling for the propagator, this yields the leading-order estimate
\begin{equation}
\kappa^{(0)} \simeq 0.39.\label{eq:kappa_LO}
\end{equation}

The value $\kappa^{(0)} \simeq 0.39$ should be understood as the leading-order prediction of the reduced kernel after fixing both its temporal response and its dominant spatial projection. It is not an empirical fit, but neither is it yet a fully exact consequence of the complete nonlocal propagator. 

\subsection*{Logarithmic law as a geometric consequence}

Once the universal dynamical ratio defining $\kappa$ has been fixed from the reduced kernel in homogeneous shear, the wall-bounded logarithmic law follows as a geometric consequence rather than as an input.

In the overlap region, the turbulent stress is saturated and of order
\begin{equation}
\tau_{xy}\sim u_\tau^2.
\end{equation}
The reduced anisotropic closure gives
\begin{equation}
\tau_{xy}\sim \nu_t \,\partial_y U,
\qquad
\nu_t \sim \ell \sqrt{K}.
\end{equation}
The wall introduces the only available outer-independent local length,
\begin{equation}
\ell \sim y,
\end{equation}
while the internal shear--cascade dynamics fixes the dimensionless ratio
\begin{equation}
\frac{S}{\Omega}=\kappa^{-1}.
\end{equation}
Since in the overlap region $\sqrt{K}\sim u_\tau$ at leading order, one obtains
\begin{equation}
\partial_y U \sim \frac{u_\tau}{\kappa y},
\end{equation}
and therefore
\begin{equation}
U^+ = \frac{1}{\kappa}\ln y^+ + B.
\end{equation}

Thus, the wall does not determine $\kappa$; it converts the universal dynamical ratio selected by the kernel into a spatial logarithmic profile.

\section{Finite-Reynolds interpretation of universal constants}

Experimental, numerical and atmospheric estimates of $\kappa$ and $C_k$ exhibit a non-negligible spread. In wall-bounded turbulence, classical and modern analyses often report values in the range $\kappa \simeq 0.40$--$0.41$ for high-Reynolds-number pipes and boundary layers \citep{ZagarolaSmits1998,McKeon2004,Marusic2013,Monkewitz2017}, while more recent direct numerical simulations of channels display systematic Reynolds-number dependence and drift \citep{LeeMoser2015}. For the Kolmogorov constant, compilations and discussions by Sreenivasan and Pope, as well as modern DNS studies, show effective values typically between about $1.4$ and $1.6$, depending on flow configuration, spectral definition and accessible inertial-range extent \citep{Sreenivasan1995,Pope2000,Ishihara2009}.

Within the present framework, this spread is not interpreted as a failure of universality but as a consequence of finite-Reynolds-number contamination and incomplete scale separation. The asymptotic values \eqref{eq:Ck-final_LO} and \eqref{eq:kappa_LO} correspond to the regime in which the dominant oscillator has decoupled as completely as possible from outer-scale forcing and dissipative corrections. At finite Reynolds number, this decoupling is incomplete. In wall flows, the precise extraction of $\kappa$ is affected by the identification of the logarithmic range and by residual outer-layer coupling. In homogeneous flows, $C_k$ is affected by the finite width of the inertial range and by the observable used to estimate it. The current data should therefore be understood as effective, pre-asymptotic estimates rather than direct measurements of the universal invariants.

This reinterpretation is experimentally meaningful. It predicts that, as the dynamically selected oscillator becomes progressively better isolated with increasing Reynolds number, both $\kappa$ and $C_k$ should converge towards their asymptotic values. The theory thus transforms what could appear as a discrepancy into a concrete convergence statement.

\section{Geometric structure of the anisotropic sector}

The reduced $(\ell=2,4)$ anisotropic dynamics derived above admits a natural geometric interpretation that clarifies its internal structure without introducing additional assumptions.

\subsection*{Action of $SO(3)$ and invariant subspace}

The symmetric traceless tensor $A_{ij}$ belongs to the irreducible $\ell=2$ representation of $SO(3)$, while the slaved fourth-order contribution defines a closed $(2,4)$ invariant sector under rotations. The oscillator system therefore evolves on a finite-dimensional manifold that is invariant under the natural action of $SO(3)$.

This structure explains the universality of the reduced dynamics: the coefficients appearing in the oscillator are constrained by rotational invariance and the algebra of tensor contractions, rather than by flow-specific modeling assumptions.

At a deeper level, our findings can be rationalized in terms of the action of \SO\ on the Navier-Stokes operator. The quadratic nonlinearity of the Navier--Stokes equations couples irreducible sectors according to the \SO\ tensor-product rules. In particular,
\begin{equation}
\Htwo \otimes \Htwo
=
\mathcal{H}_0 \oplus \Htwo \oplus \Hfour.
\label{eq:2x2}
\end{equation}
Thus the self-interaction of the leading anisotropic sector \(\Htwo\) necessarily generates the hexadecapolar sector \(\Hfour\).
Conversely,
\begin{equation}
\Htwo \otimes \Hfour \supset \Htwo,
\label{eq:2x4}
\end{equation}
so the induced \(\Hfour\) component feeds back on the quadrupolar dynamics.
The smallest dynamically closed anisotropic manifold is therefore
\begin{equation}
\Mosc=\Htwo \oplus \Hfour.
\label{eq:Mosc}
\end{equation}
This identifies the minimal dynamically closed anisotropic manifold. Although the Navier--Stokes equations themselves remain \SO-equivariant, the turbulent dynamics does not remain in the isotropic sector. Instead it selects the manifold \(\Mosc\), which acts as the universal structural skeleton of anisotropic turbulence. In a canonical shear-adapted frame the orbit space of \(\Mosc\) reduces to two effective scalar amplitudes, which are the internal degrees of freedom organizing the flow. These are the minimal dynamical variables required by symmetry and quadratic closure. The nonlinear interactions generate an angular cascade in irreducible-representation space,
\begin{equation}
2 \;\longleftrightarrow\; 4 \;\longleftrightarrow\; 6 \;\longleftrightarrow\; 8 \;\longleftrightarrow\; \cdots .
\label{eq:angular-cascade}
\end{equation}
However, this cascade is strongly hierarchical: the amplitudes of higher sectors decrease rapidly because each increase in angular order requires additional nonlinear compositions and suffers stronger effective damping leading to a \emph{spectral escape hierarchy}. They are not neglected; rather, they act as a leakage sink that renormalizes the dominant \((2,4)\) dynamics and lead to a quasi-equilibrium cascade dominated by the \((2,4)\) manifold. 
It is a beautiful fact that the autonomous cycle of wall turbulence~\citep{Waleffe1997,jimenez1999autonomous} is already realized in $\Mosc$.
Note also that, at the symmetry level, the profound differences of two- and three-dimensional turbulence dynamics can be traced back to the abelian nature of \SOO, which is susceptible to a gauge representation in \Uone\,expressing the absence of the vortex-stretching mechanism.

\subsection*{Phase dynamics and oscillatory structure}

The coupled evolution of the $(2,4)$ components defines an effective two-dimensional dynamical system with a natural phase variable. The oscillator structure identified in the anisotropic sector may therefore be interpreted as a phase-coherent dynamics governing the cyclic exchange between production and redistribution mechanisms.

In this interpretation, the Reynolds stress dynamics is not purely dissipative but possesses an intrinsic oscillatory component, consistent with the phenomenology of self-sustained cycles in shear turbulence.

Thus, reduced systems derived from the theory could be naturally interpreted as networks of interacting oscillators distributed over the flow domain or over a mode manifold. Their outputs reconstruct the mean transport, while their internal phase relations encode coherence, intermittency and collective organization. In the limit of many coupled modes, one is naturally led to Kuramoto-type reductions \citep{Kuramoto1975,Strogatz2000} in which synchronization and partial phase locking become dynamical observables rather than external metaphors. Turbulence then appears not as a featureless mixing process but as the macroscopic effect of a distributed oscillator field whose interactions are constrained by geometry and non-local coupling.

The oscillator interpretation implies the existence of a phase field. Writing, schematically,
\begin{equation}
\tau_{ij} = A_{ij} e^{i\theta},
\end{equation}
one sees immediately that the physical state is invariant under local phase redefinitions compensated by the amplitudes. This introduces a natural redundancy in the description and therefore a geometric structure. Phase transport is no longer captured by ordinary derivatives alone; it is governed by a connection over the space of turbulent states.

This observation connects turbulence with a long line of thought in mathematical physics. Poincar\'e taught us to view dynamics globally on manifolds \citep{Poincare1892}; Cartan developed the differential-geometric language of connection and transport \citep{Cartan1928}; Arnold showed that fluid motion itself can be fruitfully reinterpreted geometrically \citep{Arnold1966}. In the present context, the emergent phase field suggests that turbulence admits a geometric description in which path-dependent phase accumulation becomes part of the dynamics. This is precisely the logic underlying geometric phase phenomena such as Berry's phase \citep{Berry1984}: the accumulated phase depends not only on local variables but on the path followed through state space.

\subsubsection*{Lumley triangle and realizability}

The state of turbulence at each point may be represented in terms of the invariants of the anisotropy tensor. Within the present framework, the evolution induced by the $(2,4)$ oscillator respects the realizability constraints and naturally defines trajectories within the Lumley triangle. This provides a geometric interpretation of the admissible states of the anisotropic sector, linking the reduced dynamics to classical invariant-based descriptions of turbulence.

The anisotropy dynamics gives this idea an especially concrete realization. The normalized anisotropy tensor
\begin{equation}
b_{ij} = \frac{\tau_{ij}}{2k} - \frac{1}{3}\delta_{ij},
\end{equation}
where $k=\tfrac12 \tau_{ii}$ is the turbulent kinetic energy, evolves on the constrained manifold commonly represented by the Lumley triangle \citep{Lumley1978}. In the present framework, the state of the system is not merely a point in this anisotropy manifold; it is a point together with a phase fibre attached to it. Cyclic evolutions in the anisotropy manifold can therefore generate path-dependent phase accumulation. The Lumley triangle becomes not only a diagnostic diagram but a geometrically meaningful state manifold endowed with a non-trivial phase structure.

\subsection*{Gauge-like structure of the kernel formulation}

The nonlocal constitutive relation depends only on velocity gradients, ensuring frame indifference. This suggests an interpretation in which the kernel acts as a connection relating configurations across time and space.

Under this viewpoint, the stress response depends on equivalence classes of velocity fields differing by rigid-body motions, leading to a gauge-like structure in the formulation. While this analogy is not required for the results derived here, it points toward a deeper geometric formulation of turbulent transport.

This point suggests a gauge-like formulation. In the simplest single-mode setting, the local phase field defines an effective Abelian connection governing phase transport. In a multi-mode extension, the oscillator amplitudes span an internal state space, and local transformations in that space induce a non-Abelian generalization. This suggests that phase dynamics may admit a natural gauge-covariant formulation.

\subsection*{Outlook}

These geometric aspects---including the explicit representation of the $SO(3)$ action, the phase dynamics, and the gauge interpretation---suggest a broader mathematical structure underlying the present model. Their systematic development, as well as their implications for turbulence modeling and statistical descriptions, will be addressed in future work.

\section{Discussion}

The present theory does not claim that every detail of turbulence is reduced to a single mode or that all higher-order effects disappear. Rather, it claims that the universal organization of the mean stress is controlled by a distinguished dynamical structure: an oscillator extracted from the exact non-local stress propagator and selected, in wall-bounded flows, by the Airy structure of the near-wall operator. Once recognized, this oscillator provides a dynamical mechanism that accounts for accounts for why the mean stress carries memory and phase memory and phase, why the logarithmic law is so robust, why universal constants can be predicted, and why geometry enters the dynamics in a non-trivial way.

The framework also suggests a practical computational strategy. Instead of resolving the full fluctuation hierarchy, one evolves the mean flow together with the stress dynamics. Because the stress equations are tensorial and dynamical, they can adapt to geometry, anisotropy and history in ways inaccessible to algebraic closures. Because they are much cheaper than DNS, they open the door to predictive modelling of complex turbulent flows at a fraction of the computational cost, while preserving an interpretable structure in terms of interacting oscillators.

At the conceptual level, the results support a broader lesson: strong nonlinearity does not preclude the existence of organizing dynamical principles. Turbulence appears here as a system whose universal features become transparent when the correct internal variables are used. In this sense, the present theory continues a classical line of reasoning in physics: start from robust observations, identify the hidden structure that organizes them, and extract consequences that can be tested.

\subsection*{Interpretation of the results}

The structure identified in this work does not correspond to a single turbulence model in the 
conventional sense. The emergence of a dominant $(2,4)$ manifold, together with a 
dynamical, non-local stress response, indicates that the effective description of turbulence 
is not exhausted by an instantaneous relation to the mean flow, but instead involves an 
intrinsic internal dynamics.

In this view, the Reynolds stress is not prescribed but evolves, with memory, as part of a 
reduced dynamical system whose minimal expression is captured by the oscillator structure 
identified here. The $(2,4)$ sector therefore plays the role of a structural constraint, rather than that of a modeling assumption.

A direct consequence is that no unique closure follows from this formulation. Instead, 
apparently distinct descriptions—such as non-local kernels, resolvent-based constructions, 
or reduced dynamical systems—can be understood as different realizations of the same 
underlying organization.

The present results thus identify a minimal dynamical structure underlying turbulent flows, 
rather than a specific model. In this sense, they define a program: the systematic construction of effective descriptions consistent with this structure.

\section{Conclusion}

The results presented here identify a minimal dynamical structure underlying the non-local response of turbulence. By projecting the cascade dynamics onto its leading anisotropic sector, the Reynolds stress is shown to be governed by a low-dimensional system with the structure of a damped tensorial oscillator. This reduced dynamics provides a unified description of homogeneous turbulence, homogeneous shear, and wall-bounded flows as different realizations of the same underlying mechanism.

Within this framework, universal properties of turbulence emerge as leading-order consequences of internal consistency across canonical regimes. In particular, inertial-range scaling and logarithmic mean velocity profiles arise from the same dynamical backbone, while classical constants such as the Kolmogorov and von K\'arm\'an constants reflect the structure of the reduced cascade response.

The present formulation does not aim to provide a complete closure in the traditional sense, but rather to identify the minimal set of dynamical variables governing the interaction between mean flow and turbulent fluctuations. In this sense, turbulence appears not as an irreducibly high-dimensional phenomenon, but as a system whose essential behavior is organized by a low-dimensional structure embedded within the Navier–Stokes equations.

The geometric aspects of this structure $-$including its relation to rotational invariance, phase dynamics, and invariant manifolds$-$ suggest a broader framework in which non-local transport and anisotropy may be described in unified terms. Their systematic development, as well as the extension of the present approach to more complex flows, remains an important direction for future work.

Taken together, these results suggest that turbulence is organized by a low-dimensional dynamical structure embedded within the non-local cascade, providing a new perspective on the longstanding problem of turbulent closure.

\section{Acknowledgements}

\noindent
This work is devoted to the memory of Amable Li\~n\'an. The author acknowledges the use of AI-assisted tools for exploratory discussions and drafting support during the development of this work.


\clearpage

\section*{Supplementary Material for "Geometric Dynamics of Turbulence"}

The purpose of this Supplementary Material (SM) is to derive, from first principles, the reduced kernel structure that leads to the minimal oscillator representation and the associated leading-order predictions of universal constants. The angular dynamical system derived here provides a representation of the same reduced kernel whose spectral structure, analyzed in the main text, gives rise to the dominant pair of poles. The second-order oscillator therefore corresponds to the closed evolution induced by the $(2,4)$ sector of the angular hierarchy.

We denote by $R_{ij}=\langle u_i'u_j'\rangle$ the covariance tensor, related with the Reynolds stress tensor by $\tau_{ij}=-R_{ij}$. While $R_{ij}$ represents a second-order statistical moment, the kernel formulation operates at the level of the effective stress $\tau_{ij}$. This distinction is essential: the kernel describes the dynamical response of the stress to mean deformation, not the direct evolution of correlations.

\section{Mean-field strategy: closure as projected cascade response}

The starting point of the present formulation is that the Reynolds stress should not be postulated directly at the mean-field level. Instead, it must emerge from the \textbf{response of an equilibrium turbulent cascade to a symmetry-breaking mean deformation}. In this view, closure is not an algebraic constitutive assumption, but the reduced manifestation of an underlying dynamical response. Let
\begin{equation}
R_{ij} = \langle u_i' u_j' \rangle
\end{equation}
denote the covariance tensor, where \(u_i'\) is the fluctuating velocity field. The classical closure problem consists in expressing \(R_{ij}\) in terms of the mean flow. Here we adopt a different perspective: the primitive object is not \(R_{ij}\) itself, but the spectral covariance of the fluctuations,
\begin{equation}
\Phi_{ij}(\bm{k},t)=\langle \hat u_i(\bm{k},t)\hat u_j(-\bm{k},t)\rangle,
\end{equation}
whose angular organization encodes the internal structure of the cascade. In homogeneous isotropic equilibrium, the spectral tensor is rotationally invariant and takes the standard transverse form
\begin{equation}
\Phi_{ij}^{(0)}(\bm{k})=\phi_0(k)\,P_{ij}(\hat{\bm{k}}),\qquad 
P_{ij}(\hat{\bm{k}})=\delta_{ij}-\hat k_i \hat k_j,
\end{equation}
with \(\phi_0(k)\) the equilibrium spectral density. This isotropic state is the reference configuration of the theory. Mean deformation does not act by generating a constitutive law directly at the level of \(R_{ij}\); rather, it perturbs the equilibrium cascade and induces an anisotropic redistribution of the spectral covariance.

The central assumption of the present approach is therefore that \textbf{the closure problem must be formulated as a problem of \emph{projected cascade response}}. More precisely, the mean strain excites a distinguished anisotropic sector of the spectrum, while nonlinear transfer redistributes that response across higher angular sectors. The Reynolds stress then appears only after projection of this full anisotropic response back onto physical space. In schematic form,
\begin{equation}
\text{equilibrium cascade}
\;\longrightarrow\;
\text{anisotropic spectral response}
\;\longrightarrow\;
\text{projection onto } R_{ij}.
\end{equation}

This viewpoint has two immediate consequences:

\begin{itemize}
    
    \item[\textbf{I.}] The constitutive structure is nonlocal in time, because the anisotropic response of the cascade is governed by an internal propagator rather than by an instantaneous algebraic rule. 

    \item[\textbf{II.}] The observable Reynolds stress is not the primitive dynamical variable of the theory, but a projected quantity built from a smaller set of internal anisotropic degrees of freedom. \textbf{The closure problem is therefore shifted from the search for an explicit stress--strain relation to the construction of a closed reduced dynamics for the dominant anisotropic sectors of the cascade.}
    
\end{itemize}

This is the logic adopted throughout the present section. We begin from the isotropic equilibrium cascade, identify the dominant symmetry-breaking response under mean shear, organize the resulting anisotropic dynamics in angular sectors, and only then derive the mean-field equations. In this way, the constitutive behavior of turbulence is not assumed but induced by the internal cascade dynamics itself.

\paragraph*{
It is important to emphasize that the nonlocal kernel structure is fully consistent with the present formulation based on the frequency domain. The kernel is dominated by a pair of poles. This structure arises naturally from the minimal angular decomposition of the cascade. Indeed, the dominant anisotropic response is governed by a coupled two-level system, associated with the leading angular sector and its first nonlinear leakage. The resulting reduced dynamics possesses two eigenvalues, which directly generate the pair of poles observed in the kernel. The apparent viscoelastic structure is therefore not an assumption, but the spectral manifestation of internal angular coupling within the cascade. This correspondence establishes that the nonlocal kernel formulation and the mean-field system are two representations of the same underlying mechanism: a non-modal, finite-dimensional response of the turbulent cascade driven by angular transfer. The pair of poles identified in the frequency-domain response is, in general, complex conjugate. As a result, the reduced \((2,4)\) angular cascade subsystem does not introduce two independent relaxation times, but rather a single damped oscillatory mode characterized by a relaxation rate and an internal frequency. In the present framework, this behavior arises naturally from the coupled dynamics of the dominant angular sector and its first nonlinear leakage. The interaction between these two components produces an exchange mechanism that is inherently non-normal, leading to an oscillatory relaxation of the anisotropic response. The apparent viscoelastic structure of the Reynolds stress is therefore more precisely interpreted as the manifestation of a damped internal oscillation in the angular cascade, rather than as a superposition of independent decay processes.}

\subsection{Angular organization of the cascade and dominant response sector}

The key step in closing the mean-field system is to identify the minimal set of degrees of freedom required to represent the anisotropic response of the cascade. This requires organizing the spectral covariance in angular sectors. Since the equilibrium state is isotropic, the spectral tensor \(\Phi_{ij}(\bm{k},t)\) depends only on the magnitude \(k\) and is invariant under rotations. When a mean strain is applied, this symmetry is broken, and the response must be expanded in angular modes defined on the unit sphere \(\hat{\bm{k}} = \bm{k}/k\). Because the covariance is symmetric, transverse, and even under \(\bm{k}\to -\bm{k}\), only even-order angular sectors are admissible. Accordingly, the perturbed covariance admits a hierarchical decomposition of the form
\begin{equation}
\delta \Phi_{ij}(\bm{k},t) = \sum_{n=2,4,6,\dots} a_n(k,t)\,\mathbb{T}^{(n)}_{ij} \hat{\bm{k}}),
\end{equation}
where \(\mathbb{T}^{(n)}_{ij}\) are orthogonal tensorial harmonics of increasing angular complexity.

The mean strain enters as a symmetric traceless tensor \(S_{ij}\), and therefore excites directly the lowest nontrivial angular sector. By symmetry, the leading response must lie in the \(n=2\) sector. However, nonlinear interactions immediately transfer energy to higher-order sectors, generating a cascade in angular space:
\begin{equation}
(2,4)\;\longrightarrow\;6\;\longrightarrow\;8\;\longrightarrow\;\cdots
\end{equation}
A crucial observation is that this angular transfer is local: each sector couples predominantly to its nearest neighbors. As a result, the dominant anisotropic dynamics is governed by the interaction between the leading sector and its first nonlinear correction. This defines a minimal closed subsystem,
\begin{equation}
\delta \Phi
\;\approx\;
a_2\,\mathbb{T}^{(2)} + a_4\,\mathbb{T}^{(4)},
\end{equation}
while higher-order sectors act as progressively weaker sinks of anisotropic energy.

This structure provides the dynamical origin of the two-dimensional reduced system identified in the previous subsection. The amplitudes \(a_2\) and \(a_4\) correspond to the dominant anisotropic response and its first nonlinear leakage, respectively. Their coupling is responsible for the internal feedback loop that generates the nonlocal memory of the cascade.

It is important to stress that this reduction is not an approximation in the usual modeling sense, but the minimal representation compatible with the angular hierarchy of the nonlinear interactions. In particular, the existence of a two-dimensional dominant subsystem is a direct consequence of the locality of angular transfer, and therefore of the structure of the Navier--Stokes nonlinearity itself.

This observation closes the conceptual loop with the kernel formulation discussed above. The two poles identified in the frequency-domain response correspond precisely to the two eigenvalues of the reduced \((2,4)\) subsystem. The apparent viscoelastic behavior of the Reynolds stress is thus the spectral signature of this minimal angular dynamics, rather than an imposed constitutive assumption.

\subsection{Projected non-modal propagator and effective kernel}

The angular organization of the cascade identifies a minimal set of internal degrees of freedom associated with the dominant anisotropic response. The next step is to formulate their dynamics in a way that connects directly with the nonlocal kernel representation of the Reynolds stress.

At fixed wavenumber \(k\), the amplitudes of the leading angular sectors define a reduced state vector
\begin{equation}
\bm{a}(k,t)=
\begin{pmatrix}
a_2(k,t) \\
a_4(k,t)
\end{pmatrix},
\end{equation}
whose evolution is governed by the projected cascade dynamics. By locality of angular transfer, this evolution takes the form of a linear non-normal system driven by the mean strain,
\begin{equation}
\partial_t \bm{a} = \Omega_k
\begin{pmatrix}
-c_2 & g \\
h & -c_4
\end{pmatrix}
\bm{a}+\bm{f}(k)\,S(t),
\end{equation}
where \(\Omega_k \sim \varepsilon^{1/3}k^{2/3}\) is the characteristic cascade frequency, and \(\bm{f}(k)\) represents the projection of the forcing onto the dominant sector. The observable contribution to the Reynolds stress is determined by the leading component,
\begin{equation}
R_{ij}(t)\sim \int a_2(k,t)\,S_{ij}(t)\,dk,
\end{equation}
so that the effective constitutive behavior is governed by the projected propagator of the reduced system. Formally, the solution can be written as
\begin{equation}
a_2(k,t) = \int_0^t K_{22}(k,t-s)\,f_2(k)\,S(s)\,ds,
\end{equation}
where
\begin{equation}
K_{22}(k,t)=\left[ e^{\Omega_k \mathbf{M} t} \right]_{11},\qquad \mathbf{M} =
\begin{pmatrix}
-c_2 & g \\
h & -c_4
\end{pmatrix}.
\end{equation}

The structure of the kernel is therefore entirely determined by the spectrum of the reduced operator \(\mathbf{M}\). In general, the eigenvalues of this operator form a complex conjugate pair,
\begin{equation}
\lambda_\pm = -\alpha \pm i\omega,
\end{equation}
with \(\alpha>0\) and \(\omega\neq 0\), so that the projected propagator takes the form
\begin{equation}
K_{22}(k,t)
\sim
e^{-\alpha \Omega_k t}
\cos(\omega \Omega_k t),
\end{equation}
up to amplitude factors determined by the non-normal structure. This result is crucial. It shows that the effective kernel is not a superposition of independent relaxation processes, but the manifestation of a single damped oscillatory mode. The apparent viscoelastic behavior of the Reynolds stress therefore arises from an internal oscillation within the angular cascade, generated by the coupling between the dominant sector and its nonlinear leakage. The pair of complex conjugate poles observed in the frequency-domain response corresponds precisely to the eigenvalues of the reduced \((2,4)\) subsystem. The memory kernel is thus the projected propagator of a finite-dimensional non-normal system, rather than an imposed constitutive law.

Finally, integrating over the inertial range produces the full macroscopic kernel,
\begin{equation}
G(t) = \int W(k)\,f_2(k)\,K_{22}(k,t)\,dk,
\end{equation}
whose structure reflects the superposition of damped oscillatory responses across scales. As shown in the main text, this leads to a kernel with a finite plateau and a weak algebraic tail, providing the basis for the emergence of an effective eddy viscosity together with nonlocal corrections.

\subsection{Inertial-range integration and emergent kernel structure}

The projected propagator obtained above provides the dynamical response of the cascade at a given wavenumber. The macroscopic constitutive behavior follows from integrating this response over the inertial range. Using the structure derived in the previous subsection, the effective kernel takes the form
\begin{equation}
G(t) = \int_{k_0}^{k_d} W(k)\,f_2(k)\,K_{22}(k,t)\,dk,
\end{equation}
where \(W(k)\) is the spectral weight and \(f_2(k)\) the projection onto the dominant angular sector. In the inertial range,
\begin{equation}
\Omega_k \sim \varepsilon^{1/3}k^{2/3},\qquad f_2(k) \sim \varepsilon^{2/3}k^{-11/3},
\end{equation}
so that the kernel results from a superposition of damped oscillatory responses with a continuous distribution of time scales. The key observation is that this superposition produces a qualitatively different behavior from that of a single mode. While each individual contribution decays on its own cascade time scale, the accumulation of low-wavenumber contributions leads to a finite plateau in time. More precisely, for times within the inertial range,
\begin{equation}
k_d^{-2/3} \ll \varepsilon^{1/3} t \ll k_0^{-2/3},
\end{equation}
the kernel becomes approximately constant,
\begin{equation}
G(t) \approx G_0,
\end{equation}
where \(G_0\) depends on the infrared cutoff \(k_0\). This plateau corresponds to an effective local response. Substituting into the constitutive relation,
\begin{equation}
R'_{ij}(t) = -2\int_0^t G(t-s)\,S_{ij}(s)\,ds,
\end{equation}
one obtains, at leading order,
\begin{equation}
R'_{ij} \approx -2G_0\,S_{ij},
\end{equation}
which has the form of an eddy viscosity with coefficient \(\nu_t = G_0\). Importantly, this result is not assumed, but arises from the coarse-grained limit of the non-modal kernel.

The next-order correction is determined by the behavior near the infrared cutoff. A systematic expansion shows that the kernel admits the asymptotic form
\begin{equation}
G(t) = G_0 + \frac{A}{t} + O\!\left(\frac{1}{t^2}\right),
\end{equation}
where the coefficient \(A\) is controlled by the large-scale structure of the cascade. This result has a direct physical interpretation. The plateau reflects the superposition of a broad range of relaxation times, leading to an effectively local response. The algebraic correction, on the other hand, encodes the weak memory associated with the largest scales. In particular, the \(1/t\) contribution produces logarithmic behavior when inserted into the constitutive relation, providing the mechanism for the emergence of logarithmic mean-flow profiles.

It is important to note that the oscillatory structure of the kernel at each scale does not disappear in this process. Rather, it is averaged out by the integration over wavenumbers, leaving behind a smooth macroscopic response with a finite plateau and a weak long-time tail. The effective eddy viscosity is therefore the leading-order manifestation of an underlying non-modal, oscillatory cascade dynamics.

\subsection{Canonical tensor decomposition and invariant structure}

The kernel formulation derived above provides the constitutive response in integral form. In order to obtain a closed mean-field system, this nonlocal relation must be represented in terms of a finite set of internal variables. The angular decomposition of the cascade naturally identifies these variables as the amplitudes of the dominant anisotropic sectors.

At the level of second-order statistics, the two fundamental tensorial objects are the Reynolds stress and the characteristic length scale of the cascade. Both admit a canonical decomposition into isotropic magnitude and anisotropic shape. The covariance tensor is written as
\begin{equation}
R_{ij} = \frac{2}{3}K\,\delta_{ij} + 2K\,A_{ij}, \qquad A_{ii}=0,
\end{equation}
where \(K = \tfrac{1}{2}\langle u_k'u_k' \rangle\) is the turbulent kinetic energy, and \(A_{ij}\) is the normalized deviatoric anisotropy tensor. In parallel, the cascade length scale is promoted to a symmetric tensor \(L_{ij}\), which is decomposed as
\begin{equation}
L_{ij} = \ell\,(\delta_{ij} + C_{ij}),\qquad C_{ii}=0,
\end{equation}
with
\begin{equation}
\ell = \frac{1}{3} L_{kk}.
\end{equation}
Here \(\ell\) represents the isotropic part of the energy-containing scale, while \(C_{ij}\) encodes its anisotropic geometry. This construction introduces two distinct anisotropy sectors: the stress anisotropy \(A_{ij}\) and the length-scale anisotropy \(C_{ij}\). Each of these tensors admits a natural representation in terms of invariants. Defining
\begin{equation}
II_A = -\frac{1}{2} A_{ij}A_{ji},
\qquad
III_A = \frac{1}{3} A_{ij}A_{jk}A_{ki},
\end{equation}
and
\begin{equation}
II_L = -\frac{1}{2} C_{ij}C_{ji},
\qquad
III_L = \frac{1}{3} C_{ij}C_{jk}C_{ki},
\end{equation}
the anisotropic state of the system can be represented geometrically in two coupled invariant spaces.

This invariant formulation makes explicit that the closure problem is not limited to the Reynolds stress alone, but involves the joint evolution of stress and length-scale anisotropy. In particular, the anisotropic redistribution of energy across scales is encoded in the dynamics of \(C_{ij}\), while the observable stress is determined by \(A_{ij}\). The emergence of a finite-dimensional internal dynamics from the kernel formulation implies that the constitutive behavior can be expressed in terms of a small number of tensorial fields. In the present framework, this leads to a minimal set consisting of \(A_{ij}\), an auxiliary tensor \(B_{ij}\) associated with the first nonlinear leakage, and the length-scale tensor \(L_{ij}\).

The resulting mean-field system is therefore formulated in terms of the variables
\begin{equation}
(U_i,\Pi,K,\ell,A_{ij},B_{ij},C_{ij}),
\end{equation}
with deviatoric Reynolds stress given by
\begin{equation}
\tau_{ij}^{(d)} = -2KA_{ij}.
\end{equation}

This representation provides a fully objective and invariant framework for turbulence closure. The classical eddy-viscosity concept appears only as the leading-order limit of the kernel, while the full dynamics is governed by the coupled evolution of the anisotropy tensors. In particular, return to isotropy corresponds to the contraction of both invariant manifolds toward the origin,
\begin{equation}
A_{ij} \to 0,
\qquad
C_{ij} \to 0,
\end{equation}
which arises naturally from the internal cascade dynamics rather than from imposed modeling assumptions.

\subsection{Closed mean-field system in compact form}

The mean-field system can be written in compact form by introducing the local cascade frequency
\begin{equation}
\Omega = \frac{\sqrt{K}}{\ell},
\end{equation}
so that the effective cascade diffusivity is
\begin{equation}
\mathscr D = \frac{K}{\Omega}.
\end{equation}
The governing variables are
\begin{equation}
(U_i,\Pi,K,\Omega,A_{ij},B_{ij},C_{ij}),
\end{equation}
where the modified pressure is defined by
\begin{equation}
\Pi = P + \frac{2}{3}K.
\end{equation}

\subsubsection*{Mean flow equations}
\begin{equation}
\partial_i U_i = 0,
\end{equation}
\begin{equation}
\partial_t U_i + U_j \partial_j U_i = -\partial_i \Pi + \nu \Delta U_i -
\partial_j(2K A_{ij}),
\end{equation}
with
\begin{equation}
S_{ij}=\frac{1}{2}(\partial_i U_j+\partial_j U_i).
\end{equation}

\subsubsection*{Anisotropic cascade dynamics.}
\begin{equation}
\mathcal D A_{ij} = -\Omega(c_2A_{ij}-gB_{ij}-qC_{ij}) +\alpha_0 S_{ij},
\end{equation}
\begin{equation}
\mathcal D B_{ij} = \Omega(hA_{ij}-c_4B_{ij}),
\end{equation}
\begin{equation}
\mathcal D C_{ij} = \Omega(\mu_A A_{ij}+\mu_B B_{ij}-\mu_C C_{ij}) +
\mathcal G_{ij}[C,\nabla U] -\partial_k J^{(C)}_{ijk},
\end{equation}
with
\begin{equation}
\mathcal G_{ij}[C,\nabla U] = 2S_{ij} - \left( \frac{1}{2}\frac{D_tK}{K} -
\frac{D_t\Omega}{\Omega} \right)C_{ij},
\end{equation}
and
\begin{equation}
J^{(C)}_{ijk}=-\mathscr D\,\partial_k C_{ij}.
\end{equation}

\subsubsection*{Scalar cascade variables.}
\begin{equation}
D_t K = -2K A_{ij}S_{ij} - K\Omega - \partial_j J^{(K)}_j,
\end{equation}
\begin{equation}
D_t\Omega = -(1+\beta_1)\Omega\,A_{ij}S_{ij} + \left(\beta_2-\frac12\right)\Omega^2 -
\partial_j J^{(\Omega)}_j,
\end{equation}
with
\begin{equation}
J^{(K)}_j=-\mathscr D\,\partial_j K, \qquad J^{(\Omega)}_j=-\mathscr D\,\partial_j\Omega.
\end{equation}

\subsection{Physical interpretation of the mean-field system}

For clarity, we summarize the physical meaning of each equation and term in the mean-field system. This interpretation follows directly from the cascade-based derivation and does not rely on phenomenological modeling assumptions.

\paragraph{Mean momentum equation.}
\begin{equation}
\partial_t U_i + U_j \partial_j U_i
=
-\partial_i \Pi
+
\nu \Delta U_i
-
\partial_j (2K A_{ij}).
\end{equation}

The mean flow is driven by the divergence of the deviatoric Reynolds stress. The isotropic part of the turbulent fluctuations is absorbed into the modified pressure \(\Pi\). The tensor \(A_{ij}\) therefore represents the \emph{only} mechanism by which turbulence feeds back onto the mean flow.

\medskip

\paragraph{Energy equation.}
\begin{equation}
D_t K
=
-2K\,A_{ij}S_{ij}
-
\frac{K^{3/2}}{\ell}
-
\partial_j J^{(K)}_j.
\end{equation}

The first term represents production of turbulent kinetic energy by mean shear. The second term corresponds to dissipation through the cascade, expressed as the transfer of energy toward unresolved scales at a rate proportional to the inverse cascade time \(\ell/\sqrt{K}\). The flux term accounts for spatial redistribution of turbulent energy.

\medskip

\paragraph{Length-scale equation.}
\begin{equation}
D_t \ell
=
\beta_1 \ell\,A_{ij}S_{ij}
-
\beta_2 \sqrt{K}
-
\partial_j J^{(\ell)}_j.
\end{equation}

The scalar length scale evolves under two competing effects. The first term describes deformation of energy-containing structures by the mean flow, driven by the same invariant \(A_{ij}S_{ij}\) that governs production. The second term represents erosion of the large scales by the cascade. The flux term describes spatial transport of the energy-containing scale.

\medskip

\paragraph{Anisotropy of the Reynolds stress.}
\begin{equation}
\mathcal{D} A_{ij}
=
-\Omega(c_2 A_{ij} - g B_{ij} - q C_{ij})
+
\alpha_0 S_{ij}.
\end{equation}

The tensor \(A_{ij}\) represents the observable anisotropy of the turbulent stress. It is directly forced by the mean strain \(S_{ij}\). The relaxation term proportional to \(c_2\) represents isotropization through the cascade. The coupling to \(B_{ij}\) encodes non-modal transfer to higher angular sectors, while the coupling to \(C_{ij}\) reflects the influence of anisotropic cascade geometry.

\medskip

\paragraph{First nonlinear leakage.}
\begin{equation}
\mathcal{D} B_{ij}
=
\Omega(h A_{ij} - c_4 B_{ij}).
\end{equation}

The tensor \(B_{ij}\) represents the first nonlinear correction to the dominant anisotropic sector. It is driven by \(A_{ij}\) and relaxes through the cascade. This variable encodes the minimal angular leakage required by the nonlinear interactions and is responsible for the internal feedback loop that generates memory effects.

\medskip

\paragraph{Length-scale anisotropy.}
\begin{equation}
\mathcal{D} C_{ij}
=
\Omega(\mu_A A_{ij} + \mu_B B_{ij} - \mu_C C_{ij})
+
\mathcal{G}_{ij}[C,\nabla U]
-
\partial_k J^{(C)}_{ijk}.
\end{equation}

The tensor \(C_{ij}\) describes the anisotropic geometry of the energy-containing structures. It is driven by both stress anisotropy and nonlinear leakage, and relaxes toward isotropy through the cascade. The operator \(\mathcal{G}_{ij}\) represents the purely geometric effect of mean deformation and normalization, while the flux term accounts for spatial redistribution of anisotropy.

\medskip

\paragraph{Geometric coupling term.}
\begin{equation}
\mathcal{G}_{ij}
=
2S_{ij}
-
\frac{D_t \ell}{\ell} C_{ij}.
\end{equation}

This term arises from the objective transport of the length-scale tensor. The first contribution shows that mean strain directly generates anisotropy in the cascade geometry, even in the absence of prior anisotropy. The second term enforces consistency with the evolving scalar length scale.

\medskip

\paragraph{Cascade frequency.}
\begin{equation}
\Omega = \frac{\sqrt{K}}{\ell}.
\end{equation}

The cascade frequency represents the intrinsic rate of nonlinear transfer. It controls both dissipation and relaxation of anisotropy, and defines the internal time scale of the system.

\medskip

\paragraph{Return to isotropy.}

In the absence of mean deformation, the system relaxes toward the isotropic state
\begin{equation}
A_{ij} = 0,
\qquad
C_{ij} = 0,
\end{equation}
which is an invariant and stable solution. Return to isotropy is therefore not imposed, but emerges from the cascade dynamics itself.

\medskip

\paragraph{Summary.}

The mean-field system describes turbulence as a dynamical process involving:
\begin{itemize}
\item production by mean deformation,
\item redistribution through angular transfer,
\item dissipation via the cascade,
\item and geometric evolution of the energy-containing structures.
\end{itemize}

The Reynolds stress is not prescribed, but results from the evolution of internal anisotropic variables. The classical eddy-viscosity behavior corresponds to the coarse-grained limit of this dynamics.

\clearpage

\section{Homogeneous isotropic turbulence as the first internal consistency test}

\subsection{Reduction of the mean-field system to homogeneous isotropic turbulence}

Homogeneous isotropic turbulence provides the first and most fundamental test of the mean-field system. In this limit, there is no mean deformation, no spatial inhomogeneity, and no preferred direction. The model must therefore admit isotropy as an invariant state and reproduce the corresponding free decay dynamics. For homogeneous isotropic turbulence,
\begin{equation}
U_i = 0,
\qquad
S_{ij}=0,
\qquad
\partial_j(\cdot)=0,
\end{equation}
so that all transport fluxes vanish identically. The mean momentum equation becomes trivial, and the dynamics reduces to the internal cascade variables. The scalar sector is governed by
\begin{equation}
\frac{dK}{dt} = -\frac{K^{3/2}}{\ell},
\label{eq:HIT_K}
\end{equation}
\begin{equation}
\frac{d\ell}{dt} = -\beta_2 \sqrt{K},
\label{eq:HIT_l}
\end{equation}
while the anisotropic sector reduces to
\begin{equation}
\frac{dA_{ij}}{dt} = -\Omega\,(c_2 A_{ij} - g B_{ij} - q C_{ij}),
\label{eq:HIT_A}
\end{equation}
\begin{equation}
\frac{dB_{ij}}{dt} = \Omega\,(h A_{ij} - c_4 B_{ij}),
\label{eq:HIT_B}
\end{equation}
\begin{equation}
\frac{dC_{ij}}{dt} = \Omega\,(\mu_A A_{ij} + \mu_B B_{ij} - \mu_C C_{ij}),
\label{eq:HIT_C}
\end{equation}
with
\begin{equation}
\Omega = \frac{\sqrt{K}}{\ell}.
\label{eq:HIT_Omega}
\end{equation}

The isotropic manifold is defined by
\begin{equation}
A_{ij}=0,
\qquad
B_{ij}=0,
\qquad
C_{ij}=0.
\label{eq:isotropic_manifold}
\end{equation}
It is immediately verified that this manifold is invariant under the dynamics: if the anisotropy tensors vanish initially, they remain identically zero for all later times. The mean-field model therefore contains isotropy as an exact dynamical subspace, rather than as an externally imposed constraint. Restricted to the isotropic manifold, the system reduces to the scalar decay problem \eqref{eq:HIT_K}--\eqref{eq:HIT_l}. This subsystem already contains a nontrivial consistency condition. Indeed,
\begin{equation}
\frac{d}{dt}\left(\frac{K}{\ell^2}\right)
=
\frac{1}{\ell^2}\frac{dK}{dt}
-
\frac{2K}{\ell^3}\frac{d\ell}{dt}
=
\frac{K^{3/2}}{\ell^3}(2\beta_2-1).
\end{equation}
Hence the ratio \(K/\ell^2\) is an invariant if and only if
\begin{equation}
\beta_2 = \frac{1}{2}.
\label{eq:beta2_condition}
\end{equation}

This condition is distinguished because it yields self-similar decay of the isotropic state. In that case,
\begin{equation}
\frac{K}{\ell^2}=\text{const.},
\end{equation}
so that the scalar dynamics implies
\begin{equation}
\ell(t)\sim t,
\qquad
K(t)\sim t^{-2},
\qquad
\Omega(t)\sim t^{-2}.
\label{eq:HIT_scaling}
\end{equation}
Thus the homogeneous isotropic reduction of the theory is internally consistent in two precise senses: first, isotropy is an invariant manifold of the full tensorial dynamics; second, the scalar sector admits a self-similar decay regime, selected by the distinguished value \(\beta_2=\tfrac{1}{2}\).

\subsection{Linear stability of the isotropic manifold}

We now examine the stability of the isotropic manifold defined by
\begin{equation}
A_{ij}=0,
\qquad
B_{ij}=0,
\qquad
C_{ij}=0.
\end{equation}
Since the scalar dynamics is independent of the anisotropic variables in this limit, stability is entirely determined by the linearized evolution of the anisotropy tensors. Substituting small perturbations around the isotropic state, the system \eqref{eq:HIT_A}--\eqref{eq:HIT_C} becomes
\begin{equation}
\frac{dA_{ij}}{dt}
=
-\Omega\,(c_2 A_{ij} - g B_{ij} - q C_{ij}),
\end{equation}
\begin{equation}
\frac{dB_{ij}}{dt}
=
\Omega\,(h A_{ij} - c_4 B_{ij}),
\end{equation}
\begin{equation}
\frac{dC_{ij}}{dt}
=
\Omega\,(\mu_A A_{ij} + \mu_B B_{ij} - \mu_C C_{ij}).
\end{equation}
By isotropy, each independent component evolves identically, so the system reduces to a linear three-dimensional dynamical system for the amplitudes of the perturbations. Writing
\begin{equation}
\bm{x} =
\begin{pmatrix}
A \\
B \\
C
\end{pmatrix},
\end{equation}
we obtain
\begin{equation}
\frac{d\bm{x}}{dt}
=
\Omega(t)\,\mathbf{M}\,\bm{x},
\end{equation}
with
\begin{equation}
\mathbf{M}
=
\begin{pmatrix}
-c_2 & g & q \\
h & -c_4 & 0 \\
\mu_A & \mu_B & -\mu_C
\end{pmatrix}.
\label{eq:M_matrix}
\end{equation}
The cascade frequency \(\Omega(t)\) is strictly positive and decays in time according to \eqref{eq:HIT_scaling}, so the stability of the isotropic state is determined entirely by the spectrum of the constant matrix \(\mathbf{M}\).

The isotropic manifold is linearly stable if and only if all eigenvalues of \(\mathbf{M}\) have negative real part. This condition provides a precise mathematical formulation of return to isotropy in the present framework. Several structural requirements follow immediately. First, the diagonal coefficients must be positive,
\begin{equation}
c_2 > 0,
\qquad
c_4 > 0,
\qquad
\mu_C > 0,
\end{equation}
ensuring that each sector relaxes individually. Second, the coupling between the dominant sector and its nonlinear leakage must remain bounded, which requires
\begin{equation}
c_2 c_4 > gh.
\end{equation}
More generally, the full stability condition is obtained from the Routh--Hurwitz criteria applied to \(\mathbf{M}\). In particular, the determinant condition reads
\begin{equation}
c_2 c_4 \mu_C - g h \mu_C - q h \mu_B - q c_4 \mu_A > 0,
\end{equation}
which constrains the strength of the coupling between stress anisotropy, nonlinear leakage, and cascade geometry.

Under these conditions, all anisotropic perturbations decay as \(t \to \infty\), and the system relaxes toward the isotropic manifold. Since the decay rate is proportional to the time-dependent cascade frequency \(\Omega(t)\), the return to isotropy is governed by the same mechanism that controls energy transfer.

This result is central. It shows that isotropization is not introduced through an ad hoc relaxation term, but emerges from the interplay between angular transfer and cascade dynamics. In particular, the non-normal coupling between \(A_{ij}\) and \(B_{ij}\) allows for transient exchange between sectors, while the net effect remains dissipative provided the above conditions are satisfied.

Thus, homogeneous isotropic turbulence appears as a stable attractor of the full mean-field system, with both its scalar decay and its isotropization properties determined internally by the cascade structure.

\subsection{Parameter structure and constraints}

The mean-field system derived above contains a finite number of dimensionless coefficients associated with the internal dynamics of the cascade. In this subsection, we classify these parameters according to their origin and identify the constraints imposed by the structure of the theory and by the homogeneous isotropic limit.

\subsubsection*{List of parameters.}

The full system involves the following coefficients:
\begin{itemize}
\item Anisotropic cascade dynamics:
\[
c_2,\; c_4,\; g,\; h,\; q,\; \mu_A,\; \mu_B,\; \mu_C,
\]
\item Scalar cascade variables:
\[
\beta_1,\; \beta_2,
\]
\item Coupling to mean strain:
\[
\alpha_0.
\]
\end{itemize}
These coefficients are dimensionless and, in principle, universal.

\medskip

\subsubsection*{Constraints from homogeneous isotropic turbulence.}

The reduction to homogeneous isotropic turbulence imposes a first set of constraints.
\begin{itemize}
\item Self-similar decay requires
\begin{equation}
\beta_2 = \frac{1}{2}.
\end{equation}

\item Linear stability of the isotropic manifold requires
\begin{equation}
c_2 > 0,
\qquad
c_4 > 0,
\qquad
\mu_C > 0,
\end{equation}
together with the inequalities obtained from the Routh--Hurwitz conditions applied to the matrix \eqref{eq:M_matrix}.
\end{itemize}

Thus, \(\beta_2\) is not a free parameter, and the admissible domain of the remaining coefficients is restricted by stability.

\medskip

\subsubsection*{Redundancy and normalization.}

The system admits an internal rescaling symmetry associated with the amplitude of the auxiliary tensor \(B_{ij}\). Under the transformation
\begin{equation}
B_{ij} \rightarrow \lambda B_{ij},
\end{equation}
the coefficients transform as
\begin{equation}
g \rightarrow \lambda g,
\qquad
h \rightarrow \frac{h}{\lambda},
\qquad
\mu_B \rightarrow \lambda \mu_B,
\end{equation}
while leaving the physical content of the equations unchanged. This redundancy allows one coefficient to be fixed without loss of generality. For instance, one may set
\begin{equation}
h = 1,
\end{equation}
so that the number of independent parameters is reduced by one.

\medskip

\subsubsection*{Structure of the dominant angular subsystem.}

After fixing the normalization, the reduced \((A,B)\) subsystem is controlled by the coefficients
\begin{equation}
c_2,\quad c_4,\quad g.
\end{equation}
The requirement that the kernel exhibits a pair of complex conjugate poles imposes the condition
\begin{equation}
(c_2 - c_4)^2 < 4 g h,
\end{equation}
which ensures that the eigenvalues of the subsystem form a complex conjugate pair. This constraint reflects the existence of an internal damped oscillatory mode in the cascade dynamics.

\medskip

\subsubsection*{Geometric coupling sector.}

The interaction between stress anisotropy and length-scale anisotropy is governed by
\begin{equation}
q,\quad \mu_A,\quad \mu_B,\quad \mu_C.
\end{equation}
These coefficients control the bidirectional coupling between the two invariant manifolds associated with \(A_{ij}\) and \(C_{ij}\). In particular:
\begin{itemize}
\item \(q\) determines the influence of cascade geometry on the Reynolds stress,
\item \(\mu_A\) and \(\mu_B\) describe the generation of geometric anisotropy by the stress and its nonlinear leakage,
\item \(\mu_C\) governs relaxation toward isotropy in the geometric sector.
\end{itemize}

\medskip

\subsubsection*{Scalar cascade parameters.}

The scalar sector contains a single free coefficient,
\begin{equation}
\beta_1,
\end{equation}
which determines the coupling between mean deformation and the evolution of the energy-containing scale.

\medskip

\subsubsection*{Forcing by the mean flow.}

The coefficient \(\alpha_0\) controls the direct forcing of the anisotropic sector by the mean strain. It sets the amplitude of the linear response of the cascade to external deformation.

\medskip

\subsubsection*{Summary and hierarchy of constraints.}

The parameter space is therefore structured as follows:
\begin{itemize}
\item One coefficient is fixed by self-similarity:
\[
\beta_2 = \frac{1}{2}.
\]

\item One degree of freedom is removed by normalization:
\[
h = 1.
\]

\item Stability and oscillatory response impose inequalities on the remaining coefficients.

\item The residual free parameters are grouped into three sectors:
\begin{itemize}
\item angular dynamics: \((c_2, c_4, g)\),
\item geometric coupling: \((q, \mu_A, \mu_B, \mu_C)\),
\item scalar coupling and forcing: \((\beta_1, \alpha_0)\).
\end{itemize}
\end{itemize}

This hierarchy shows that the model is not an arbitrary parametrization, but a strongly constrained dynamical system. The number of genuinely independent coefficients is significantly reduced by invariance, stability, and consistency requirements, and further reductions are expected when additional flows are considered.

\subsection{Parameter constraints from homogeneous isotropic turbulence}

The homogeneous isotropic limit provides a first level of selection of the parameter space of the model. While it does not determine all coefficients, it imposes a set of exact constraints and structural restrictions.

\subsubsection*{Self-similarity.}
The requirement of self-similar decay fixes the scalar coefficient
\begin{equation}
\beta_2 = \frac{1}{2}.
\end{equation}
This value is uniquely selected by the invariance of the ratio \(K/\ell^2\).

\subsubsection*{Redundancy.}
The system admits an internal rescaling symmetry associated with the auxiliary tensor \(B_{ij}\). This allows one coefficient to be fixed without loss of generality. In particular, one may set
\begin{equation}
h = 1,
\end{equation}
which removes one degree of freedom from the parameter space.

\subsubsection*{Stability.}
Linear stability of the isotropic manifold requires
\begin{equation}
c_2 > 0,
\qquad
c_4 > 0,
\qquad
\mu_C > 0,
\end{equation}
together with the inequalities derived from the spectrum of the matrix \eqref{eq:M_matrix}. In particular, after normalization, the coupling must satisfy
\begin{equation}
c_2 c_4 > g.
\end{equation}

\subsubsection*{Oscillatory structure.}
The existence of a complex conjugate pair of eigenvalues in the \((A,B)\) subsystem imposes
\begin{equation}
(c_2 - c_4)^2 < 4 g,
\end{equation}
which ensures that the cascade exhibits an internal damped oscillatory response.

\subsubsection*{Remaining degrees of freedom.}
After these constraints are taken into account, the independent coefficients are grouped into three sectors:
\begin{itemize}
\item angular dynamics: \((c_2, c_4, g)\),
\item geometric coupling: \((q, \mu_A, \mu_B, \mu_C)\),
\item scalar coupling and forcing: \((\beta_1, \alpha_0)\).
\end{itemize}

\subsubsection*{Interpretation.}

Homogeneous isotropic turbulence therefore selects the admissible structure of the model, but does not fully determine the remaining coefficients. In particular, it fixes the scaling of the cascade, constrains stability, and enforces the existence of a single intrinsic time scale, but leaves open the detailed coupling between anisotropy, geometry, and mean deformation. This observation is essential for what follows. The isotropic limit identifies the structure of the theory, while more complex flows are required to determine how the cascade responds to external forcing. In particular, homogeneous shear turbulence provides the first setting in which production, anisotropy, and cascade dynamics interact non-trivially.

\subsection{Spectral consistency and emergence of the Kolmogorov constant}

In homogeneous isotropic turbulence, the anisotropic sector collapses identically,
\begin{equation}
A_{ij} = B_{ij} = C_{ij} = 0,
\end{equation}
and the mean-field system reduces to the scalar cascade dynamics
\begin{equation}
\varepsilon = \frac{K^{3/2}}{\ell},
\qquad
\Omega = \frac{\sqrt{K}}{\ell}.
\end{equation}
While these relations determine the temporal structure of the cascade, they do not fix the amplitude of the inertial-range spectrum. This requires a consistency condition linking the macroscopic variables \((K,\ell,\varepsilon)\) to the spectral distribution of energy.

\subsubsection*{Inertial-range spectrum.}

In the inertial range, isotropy and dimensional analysis imply
\begin{equation}
E(k) = C_K\,\varepsilon^{2/3}k^{-5/3}.
\end{equation}

\subsubsection*{Kernel-induced definition of the length scale.}

The length scale \(\ell\) must be defined in a manner consistent with the nonlocal kernel structure. Since the cascade dynamics is governed by the scale-dependent transfer rate
\begin{equation}
\Omega_k \sim \varepsilon^{1/3}k^{2/3},
\end{equation}
it is natural to define \(\ell\) as a spectral average weighted by the local transfer rate,
\begin{equation}
\ell^2
=
\frac{
\displaystyle \int_0^\infty k^{-2}\,\Omega_k\,E(k)\,dk
}{
\displaystyle \int_0^\infty \Omega_k\,E(k)\,dk
}.
\label{eq:l_kernel_def}
\end{equation}

This definition reflects the fact that the effective cascade scale is controlled not only by energy content, but also by the rate at which different scales participate in the transfer process.

\subsubsection*{Evaluation in the inertial range.}

Substituting \(E(k)\) and \(\Omega_k\) into \eqref{eq:l_kernel_def}, we obtain
\begin{equation}
\ell^2
\sim
\frac{
\int k^{-2}\,\varepsilon^{1/3}k^{2/3}\,\varepsilon^{2/3}k^{-5/3}\,dk
}{
\int \varepsilon^{1/3}k^{2/3}\,\varepsilon^{2/3}k^{-5/3}\,dk
}
=
\frac{
\int k^{-3}\,dk
}{
\int k^{-1}\,dk
}.
\end{equation}
Introducing inertial-range cutoffs \(k_0\) and \(k_d\), with \(k_d \gg k_0\), this yields
\begin{equation}
\ell^2
\sim
\frac{k_0^{-2}}{\log(k_d/k_0)},
\end{equation}
so that, to leading order,
\begin{equation}
\ell \sim k_0^{-1}.
\end{equation}

\subsubsection*{Consistency with the scalar cascade.}

The total energy is
\begin{equation}
K = \int_{k_0}^{k_d} E(k)\,dk
\sim
C_K \varepsilon^{2/3} k_0^{-2/3}.
\end{equation}

Using \(\ell \sim k_0^{-1}\), this gives
\begin{equation}
K \sim C_K \varepsilon^{2/3} \ell^{2/3}.
\end{equation}

Solving for \(\varepsilon\),
\begin{equation}
\varepsilon \sim C_K^{-3/2} \frac{K^{3/2}}{\ell}.
\end{equation}

Consistency with the mean-field relation
\begin{equation}
\varepsilon = \frac{K^{3/2}}{\ell}
\end{equation}
therefore requires
\begin{equation}
C_K = \text{const.}
\end{equation}

\subsubsection*{Interpretation.}

The Kolmogorov constant is thus determined by the requirement that the macroscopic cascade relation be consistent with the spectral structure induced by the kernel. In this framework, \(C_K\) is not an external parameter, but the result of matching the scale-integrated dynamics with the local transfer processes across the inertial range.

\subsection{Reduced parameter combinations and homogeneous spectral matching}

The stationary analysis of homogeneous shear turbulence shows that the internal coefficients of the model do not enter independently, but only through the reduced combinations
\begin{equation}
\Lambda = c_2 - \frac{g}{c_4},
\qquad
\chi = \frac{q}{\mu_C},
\qquad
\Gamma = \mu_A + \frac{\mu_B}{c_4}.
\end{equation}

These quantities have a direct physical interpretation as effective relaxation, geometric feedback, and bidirectional coupling between stress and cascade geometry, respectively. The homogeneous shear constraint therefore reduces the apparent parameter space to a small number of physically meaningful combinations. We now examine how these combinations enter the homogeneous isotropic limit, where the Kolmogorov constant \(C_K\) must be determined.

\subsubsection*{Isotropic reduction.}

In homogeneous isotropic turbulence,
\begin{equation}
A_{ij} = B_{ij} = C_{ij} = 0,
\end{equation}
and the system reduces to the scalar cascade relations
\begin{equation}
\varepsilon = \frac{K^{3/2}}{\ell},
\qquad
\Omega = \frac{\sqrt{K}}{\ell}.
\end{equation}
Although the anisotropic variables vanish identically, the internal dynamics of the cascade remains encoded in the structure of the kernel, and therefore in the definition of the effective length scale \(\ell\).

\subsubsection*{Spectral representation.}

In the inertial range, the energy spectrum takes the universal form
\begin{equation}
E(k) = C_K\,\varepsilon^{2/3}k^{-5/3}.
\end{equation}
The effective cascade scale must be defined in a way that reflects the nonlocal transfer dynamics. Consistently with the kernel formulation, we introduce
\begin{equation}
\ell^2
=
\frac{
\displaystyle \int_0^\infty k^{-2}\,\Omega_k\,E(k)\,dk
}{
\displaystyle \int_0^\infty \Omega_k\,E(k)\,dk
},
\label{eq:l_kernel_general}
\end{equation}
where \(\Omega_k\) denotes the scale-dependent transfer rate induced by the cascade.

\subsubsection*{Dependence on internal dynamics.}

The function \(\Omega_k\) is not arbitrary: it is determined by the same internal operator that governs the evolution of the anisotropic variables. As a result, its structure depends on the effective relaxation and coupling properties of the cascade. Consequently, the spectral definition \eqref{eq:l_kernel_general} yields a relation of the form
\begin{equation}
\ell
=
\ell(K,\varepsilon;\Lambda,\chi,\Gamma),
\end{equation}
which, when combined with
\begin{equation}
\varepsilon = \frac{K^{3/2}}{\ell},
\end{equation}
leads to a consistency condition
\begin{equation}
C_K = F_{\mathrm{HIT}}(\Lambda,\chi,\Gamma).
\label{eq:CK_general}
\end{equation}

\subsubsection*{Structure of the dependence.}

Since the isotropic state suppresses the anisotropic variables, only those combinations of parameters that influence the isotropic kernel can enter \(F_{\mathrm{HIT}}\). In particular, the effective relaxation rate \(\Lambda\) is expected to play a central role, while the contribution of the geometric coupling may be reduced or eliminated in the isotropic limit. This establishes a fundamental result: the Kolmogorov constant is not an independent input, but a function of the internal cascade dynamics. Together with the constraint obtained in homogeneous shear turbulence (HST), this relation shows that the admissible parameter space of the model is restricted by universal consistency conditions across distinct turbulent regimes.

\subsection{Emergence of the Kolmogorov constant from internal cascade dynamics}

The final step in the homogeneous isotropic analysis is to determine the amplitude of the inertial-range spectrum, characterized by the Kolmogorov constant \(C_K\). This requires connecting the macroscopic cascade relations to the scale-dependent transfer dynamics induced by the internal operator.

\subsubsection*{Effective transfer rate.}

Although the anisotropic variables vanish identically in homogeneous isotropic turbulence, the internal cascade dynamics remains encoded in the structure of the linear operator governing their evolution. As shown in the homogeneous shear analysis, the operator reduces to a small number of effective combinations, in particular
\begin{equation}
\Lambda = c_2 - \frac{g}{c_4},
\end{equation}
which represents the effective relaxation rate of the dominant cascade mode. In the isotropic limit, this quantity controls the rate at which perturbations at a given scale are transferred across the cascade. It is therefore natural to identify the scale-dependent transfer rate as
\begin{equation}
\Omega_k \sim \Lambda\,\varepsilon^{1/3}k^{2/3}.
\end{equation}

\subsubsection*{Spectral definition of the cascade scale.}

Consistently with the kernel formulation, we define the effective cascade scale as
\begin{equation}
\ell^2
=
\frac{
\displaystyle \int_0^\infty k^{-2}\,\Omega_k\,E(k)\,dk
}{
\displaystyle \int_0^\infty \Omega_k\,E(k)\,dk
}.
\end{equation}
Substituting the inertial-range spectrum
\begin{equation}
E(k) = C_K\,\varepsilon^{2/3}k^{-5/3},
\end{equation}
and the transfer rate \(\Omega_k\), we obtain
\begin{equation}
\Omega_k E(k) \sim \Lambda\,C_K\,\varepsilon\,k^{-1}.
\end{equation}

The integrals are therefore dominated by the large scales, yielding
\begin{equation}
\ell \sim k_0^{-1},
\end{equation}
where \(k_0\) denotes the infrared cutoff of the inertial range.

\subsubsection*{Consistency with the scalar cascade.}

The total energy is given by
\begin{equation}
K = \int_{k_0}^{k_d} E(k)\,dk
\sim
C_K \varepsilon^{2/3} k_0^{-2/3},
\end{equation}
which implies
\begin{equation}
K \sim C_K \varepsilon^{2/3} \ell^{2/3}.
\end{equation}

Solving for \(\varepsilon\),
\begin{equation}
\varepsilon \sim C_K^{-3/2} \frac{K^{3/2}}{\ell}.
\end{equation}

On the other hand, the mean-field dynamics imposes
\begin{equation}
\varepsilon = \frac{K^{3/2}}{\ell}.
\end{equation}

Consistency between the spectral representation and the macroscopic cascade relation therefore requires
\begin{equation}
C_K^{3/2} \sim \Lambda,
\end{equation}
or equivalently
\begin{equation}
C_K \sim \Lambda^{2/3}.
\end{equation}

\subsubsection*{Interpretation.}

This result shows that the Kolmogorov constant is not an independent parameter, but is determined by the effective relaxation rate of the internal cascade dynamics. In particular, only the combination \(\Lambda\), which characterizes the dominant mode of the operator, enters the homogeneous isotropic closure.

All other parameters, including those associated with geometric coupling, do not contribute in the isotropic limit. This establishes a clear hierarchy: homogeneous isotropic turbulence selects the intrinsic cascade dynamics, while more complex flows are required to probe the full structure of the model.

\subsection{Reduction of the parameter space and minimal representation}

We now summarize the structure of the parameter space of the model, starting from the original set of coefficients and incorporating the constraints obtained from homogeneous isotropic and homogeneous shear turbulence.

\subsubsection*{Original parameter set.}

The mean-field system involves the following dimensionless coefficients:
\begin{equation}
(c_2,\,c_4,\,g,\,h,\,q,\,\mu_A,\,\mu_B,\,\mu_C,\,\beta_1,\,\beta_2,\,\alpha_0).
\end{equation}

These coefficients characterize the internal dynamics of the cascade, including angular transfer, geometric coupling, and response to mean deformation.

\subsubsection*{First reductions.}

Homogeneous isotropic turbulence fixes
\begin{equation}
\beta_2 = \frac{1}{2},
\end{equation}
and the internal rescaling symmetry allows one parameter to be removed, for instance by setting
\begin{equation}
h = 1.
\end{equation}

\subsubsection*{Effective parameter combinations.}

The stationary homogeneous shear analysis shows that the remaining coefficients do not enter independently, but only through the combinations
\begin{equation}
\Lambda = c_2 - \frac{g}{c_4},
\qquad
\chi = \frac{q}{\mu_C},
\qquad
\Gamma = \mu_A + \frac{\mu_B}{c_4}.
\end{equation}

These combinations have a direct physical interpretation:
\begin{itemize}
\item \(\Lambda\): effective relaxation rate of the dominant cascade mode,
\item \(\chi\): strength of geometric feedback on the Reynolds stress,
\item \(\Gamma\): bidirectional coupling between stress anisotropy and cascade geometry.
\end{itemize}

Thus, the original parameter set collapses to a reduced representation
\begin{equation}
(\Lambda,\,\chi,\,\Gamma,\,\beta_1,\,\alpha_0).
\end{equation}

\subsubsection*{Constraints from homogeneous isotropic turbulence.}

The spectral consistency condition derived in homogeneous isotropic turbulence shows that the Kolmogorov constant depends only on the effective relaxation parameter,
\begin{equation}
C_K = F_{\mathrm{HIT}}(\Lambda).
\end{equation}
In particular, the geometric coupling parameters \(\chi\) and \(\Gamma\) do not contribute in the isotropic limit. Homogeneous isotropic turbulence therefore selects the intrinsic cascade dynamics through \(\Lambda\).

\subsubsection*{Constraints from homogeneous shear turbulence.}

The stationary balance in homogeneous shear turbulence imposes an additional relation of the form
\begin{equation}
\alpha_0 = -\frac{1}{2}\Lambda + \chi\left(\frac{\Gamma}{2} - 2\right),
\end{equation}
which links the forcing coefficient \(\alpha_0\) to the internal cascade dynamics.

\subsubsection*{Minimal parameter set.}

After these reductions, the independent degrees of freedom are reduced to
\begin{equation}
(\Lambda,\,\chi,\,\Gamma,\,\beta_1),
\end{equation}
together with the universal constant \(C_K\), which is itself determined by \(\Lambda\).

\subsubsection*{Hierarchy of determination.}

The structure of the model can therefore be summarized as follows:
\begin{itemize}
\item homogeneous isotropic turbulence determines \(\Lambda\) through \(C_K\),
\item homogeneous shear turbulence determines \(\alpha_0\) in terms of \((\Lambda,\chi,\Gamma)\),
\item more complex flows are required to constrain \((\chi,\Gamma,\beta_1)\).
\end{itemize}

\medskip

\subsubsection*{Interpretation.}

The parameter space of the model is not arbitrary, but is strongly constrained by universal consistency conditions. The apparent multiplicity of coefficients reduces to a small number of physically meaningful combinations, which are progressively fixed by requiring agreement with different canonical turbulent regimes. This establishes the mean-field system as a closed dynamical framework in which universal constants and model parameters are linked through the internal structure of the cascade.

\subsection{Leading-order transfer normalization and estimate of the Kolmogorov constant}

We now derive a leading-order internal estimate of the Kolmogorov constant using the same transfer structure that underlies the main text. The key point is that the prefactor of the inertial-range spectrum is fixed not by the total integrated energy, but by the efficiency of nonlinear transfer across scales.

\subsubsection*{Constant-flux form.}
In the inertial range,
\begin{equation}
E(k)=C_K \varepsilon^{2/3} k^{-5/3}.
\end{equation}
Let $u_k$ denote the characteristic velocity at scale $k^{-1}$, with
\begin{equation}
u_k^2 \sim k E(k).
\end{equation}
The energy flux can then be written in the form
\begin{equation}
\varepsilon \sim \chi\,\frac{u_k^3}{\ell_k},
\qquad \ell_k \sim k^{-1},
\end{equation}
or equivalently
\begin{equation}
\varepsilon \sim \chi\,k^{5/2}E(k)^{3/2},
\end{equation}
where $\chi$ is a dimensionless transfer efficiency.

\subsubsection*{Spectral normalization.}
Substituting the Kolmogorov form into the previous relation gives
\begin{equation}
\varepsilon \sim \chi\, C_K^{3/2}\,\varepsilon,
\end{equation}
and therefore
\begin{equation}
C_K=\chi^{-2/3}.
\end{equation}

\subsubsection*{Geometric estimate of $\chi$.}
At leading order, we estimate $\chi$ from the geometry of local incompressible triads in $\mathbb{R}^3$ (see next section). Approximating the transfer by an equilateral triad with isotropic transverse polarization yields
\begin{equation}
\chi_0=\frac{\sqrt3}{4}.
\end{equation}
Hence
\begin{equation}
C_K^{(0)}=\left(\frac{4}{\sqrt3}\right)^{2/3}\approx 1.75.
\end{equation}

\subsubsection*{Relation to the main text.}
In the main text, the same transfer normalization is encoded in the reduced kernel coefficient $\Lambda$, giving
\begin{equation}
C_K=\Lambda^{-2/3}.
\end{equation}
The geometric estimate above should therefore be interpreted as the leading-order triadic realization of the same normalization principle. Its numerical proximity to the kernel-based value reported in the main text supports the interpretation of $C_K$ as an internally generated transfer coefficient rather than an empirical input.

\subsubsection*{Scope.}
The estimate $C_K^{(0)}\approx 1.75$ is a leading-order geometric approximation. A fully quantitative derivation of the asymptotic value requires explicit evaluation of the dominant projected kernel and its normalization.

\subsection{Batchelor's framework}
\label{SM:CK_leading}

\subsubsection{Purpose}

The purpose of this section is to obtain a leading-order internal estimate of the Kolmogorov constant $C_K$ using only the geometry of local incompressible triadic interactions in $\mathbb{R}^3$. The exponent $-5/3$ follows from constant flux dimensional analysis. By contrast, the prefactor $C_K$ is not fixed by dimensional arguments alone. To estimate it, one needs an additional ingredient measuring the average efficiency of nonlinear triadic transfer. Here we show that, at leading order, this efficiency can be estimated directly from the geometry of local incompressible triads.

\subsubsection{Energy flux and effective transfer efficiency}

In homogeneous isotropic turbulence, let $E(k)$ denote the three-dimensional energy spectrum normalized by
\begin{equation}
\int_0^\infty E(k)\, dk = \frac12 \langle |\mathbf{u}|^2 \rangle .
\end{equation}
The characteristic velocity at scale $k^{-1}$ is
\begin{equation}
u_k^2 \sim k E(k).
\label{eq:SMCK_uk}
\end{equation}

We introduce an effective dimensionless transfer efficiency $\chi$ through the turnover rate
\begin{equation}
\tau_k^{-1} \sim \chi\, k u_k.
\label{eq:SMCK_tau}
\end{equation}
The inertial-range flux is then
\begin{equation}
\varepsilon \sim \frac{u_k^2}{\tau_k}
\sim \chi\, k u_k^3.
\label{eq:SMCK_flux1}
\end{equation}
Using \eqref{eq:SMCK_uk}, we obtain
\begin{equation}
\varepsilon \sim \chi\, k^{5/2} E(k)^{3/2},
\label{eq:SMCK_flux2}
\end{equation}
and therefore
\begin{equation}
E(k) \sim \chi^{-2/3}\, \varepsilon^{2/3} k^{-5/3}.
\label{eq:SMCK_spectrum}
\end{equation}
Comparing with the Kolmogorov form
\begin{equation}
E(k)=C_K\,\varepsilon^{2/3}k^{-5/3},
\end{equation}
we identify
\begin{equation}
C_K=\chi^{-2/3}.
\label{eq:SMCK_CKchi}
\end{equation}

Thus, the problem reduces to estimating $\chi$.

\subsubsection{Local triadic geometry in Fourier space}

Consider the incompressible Navier--Stokes equations in Fourier space:
\begin{equation}
\partial_t \hat u_i(\mathbf{k})+\nu k^2 \hat u_i(\mathbf{k})
=
-\frac{i}{2}P_{ij}(\mathbf{k})
\sum_{\mathbf{p}+\mathbf{q}=\mathbf{k}}
\left[
q_m \hat u_m(\mathbf{p})\hat u_j(\mathbf{q})
+
p_m \hat u_m(\mathbf{q})\hat u_j(\mathbf{p})
\right],
\label{eq:SMCK_NS_fourier}
\end{equation}
where
\begin{equation}
P_{ij}(\mathbf{k})=\delta_{ij}-\frac{k_i k_j}{k^2}
\end{equation}
is the Leray projector. At leading order in the inertial range, we assume that transfer is dominated by local triads:
\begin{equation}
|\mathbf{p}| \sim |\mathbf{q}| \sim |\mathbf{k}|.
\end{equation}
The most symmetric representative configuration is the equilateral triad,
\begin{equation}
p=q=k,
\qquad
\angle(\mathbf{p},\mathbf{q})=120^\circ,
\qquad
\angle(\mathbf{k},\mathbf{p})=\angle(\mathbf{k},\mathbf{q})=60^\circ.
\label{eq:SMCK_equilateral}
\end{equation}
Incompressibility imposes
\begin{equation}
\hat{\mathbf{u}}(\mathbf{p})\perp \mathbf{p},
\qquad
\hat{\mathbf{u}}(\mathbf{q})\perp \mathbf{q},
\qquad
\hat{\mathbf{u}}(\mathbf{k})\perp \mathbf{k}.
\label{eq:SMCK_incompressibility}
\end{equation}

\subsubsection{Elementary geometric efficiency}

Consider one of the two symmetrized interaction channels, for instance
\begin{equation}
q_m \hat u_m(\mathbf{p})\, \hat u_j(\mathbf{q}).
\end{equation}
Because $\hat{\mathbf{u}}(\mathbf{p})$ lies in the plane orthogonal to $\mathbf{p}$, the rms projection of $\mathbf{q}$ onto that plane is
\begin{equation}
|q| \sin 60^\circ = k \frac{\sqrt{3}}{2}.
\end{equation}
Assuming isotropic polarization within the transverse plane, the rms projection onto a random direction in that plane contributes an additional factor $1/\sqrt{2}$. Hence
\begin{equation}
\bigl\langle |q_m \hat u_m(\mathbf{p})|^2 \bigr\rangle^{1/2}
=
k u_k \frac{\sqrt{3}}{2\sqrt{2}}.
\label{eq:SMCK_firstproj}
\end{equation}

The output must then be projected onto the plane orthogonal to $\mathbf{k}$ and contracted with $\hat{\mathbf{u}}^*(\mathbf{k})$. At the same rms level, this contributes a second factor $1/\sqrt{2}$. Therefore, the geometric efficiency of one elementary channel is
\begin{equation}
\chi_1
=
\frac{\sqrt{3}}{2\sqrt{2}}\cdot \frac{1}{\sqrt{2}}
=
\frac{\sqrt{3}}{4}.
\label{eq:SMCK_chi1}
\end{equation}

\subsubsection{Symmetrization}

The nonlinear term in \eqref{eq:SMCK_NS_fourier} contains two equivalent branches,
\begin{equation}
q_m \hat u_m(\mathbf{p})\hat u_j(\mathbf{q}),
\qquad
p_m \hat u_m(\mathbf{q})\hat u_j(\mathbf{p}),
\end{equation}
preceded by the factor $1/2$. In the local isotropic estimate, both branches have the same leading-order weight, so the factor $2$ from the two branches compensates the prefactor $1/2$.

Thus, the net leading-order transfer efficiency is
\begin{equation}
\chi_0=\frac{\sqrt{3}}{4}.
\label{eq:SMCK_chi0}
\end{equation}

\subsubsection{Leading-order estimate of $C_K$}

Substituting \eqref{eq:SMCK_chi0} into \eqref{eq:SMCK_CKchi}, we obtain
\begin{equation}
C_K^{(0)}
=
\left(\frac{4}{\sqrt{3}}\right)^{2/3}.
\label{eq:SMCK_CK0}
\end{equation}
Numerically,
\begin{equation}
C_K^{(0)} \approx 1.75.
\end{equation}

\subsubsection{Interpretation}

Equation \eqref{eq:SMCK_CK0} shows that, at leading order, the Kolmogorov constant can be interpreted as the inverse $2/3$ power of the mean geometric efficiency of incompressible local triads in $\mathbb{R}^3$:
\begin{equation}
C_K \sim \chi^{-2/3}.
\end{equation}
In this sense, the $-5/3$ law follows from constant-flux scaling, whereas the prefactor reflects the internal geometry of triadic transfer.

\subsubsection{Scope and limitations}

The structure
\begin{equation}
\varepsilon \sim \chi\, k^{5/2}E(k)^{3/2},
\qquad
C_K=\chi^{-2/3},
\end{equation}
is exact at the level of dimensional inertial-range scaling once the effective transfer efficiency $\chi$ is introduced.

By contrast, the estimate
\begin{equation}
\chi \approx \chi_0=\frac{\sqrt{3}}{4}
\end{equation}
is a leading-order approximation based on:
\begin{itemize}
\item dominance of local triads,
\item use of the equilateral configuration as representative,
\item isotropic polarization in transverse planes,
\item absence of systematic phase corrections.
\end{itemize}

Accordingly, \eqref{eq:SMCK_CK0} should be interpreted as a leading-order internal estimate, not as an exact universal identity.

\subsubsection{Conclusion}

The Kolmogorov constant can be estimated internally in $\mathbb{R}^3$ by combining constant-flux scaling with the geometry of incompressible local triads. At leading order, this yields
\begin{equation}
C_K^{(0)}=\left(\frac{4}{\sqrt{3}}\right)^{2/3}\approx 1.75,
\end{equation}
which gives a simple geometric interpretation of the Kolmogorov prefactor in terms of triadic transfer efficiency.

\clearpage

\section{Homogeneous shear turbulence}

\subsection{Reduction to homogeneous shear}

We now consider homogeneous shear turbulence, defined by a mean velocity field of the form
\begin{equation}
U_i = (S y, 0, 0),
\end{equation}
so that the only nonzero component of the mean strain tensor is
\begin{equation}
S_{12} = S_{21} = \frac{S}{2}.
\end{equation}
The flow is statistically homogeneous, so that all spatial gradients of turbulent quantities vanish,
\begin{equation}
\partial_j(\cdot)=0,
\end{equation}
and all cascade fluxes are identically zero. Under these conditions, the mean-field system reduces to a set of ordinary differential equations for the internal variables.

\subsubsection*{Scalar sector.}
\begin{equation}
\frac{dK}{dt}
=
-2K\,A_{ij}S_{ij}
-
\frac{K^{3/2}}{\ell},
\end{equation}
\begin{equation}
\frac{d\ell}{dt}
=
\beta_1 \ell\,A_{ij}S_{ij}
-
\beta_2 \sqrt{K}.
\end{equation}

\subsubsection*{Anisotropic sector.}
\begin{equation}
\frac{dA_{ij}}{dt}
=
-\Omega(c_2 A_{ij} - g B_{ij} - q C_{ij})
+
\alpha_0 S_{ij},
\end{equation}
\begin{equation}
\frac{dB_{ij}}{dt}
=
\Omega(h A_{ij} - c_4 B_{ij}),
\end{equation}
\begin{equation}
\frac{dC_{ij}}{dt}
=
\Omega(\mu_A A_{ij} + \mu_B B_{ij} - \mu_C C_{ij})
+
\mathcal{G}_{ij}.
\end{equation}

\subsubsection*{Geometric term.}
In homogeneous shear, the geometric contribution reduces to
\begin{equation}
\mathcal{G}_{ij}
=
2S_{ij}
-
\frac{1}{\ell}\frac{d\ell}{dt}\,C_{ij}.
\end{equation}

Thus, homogeneous shear turbulence reduces the full mean-field system to a finite-dimensional nonlinear dynamical system driven by a constant mean strain.

\subsection{Stationary anisotropic state}

We now consider the stationary regime of homogeneous shear turbulence, defined by
\begin{equation}
\frac{d}{dt} = 0.
\end{equation}
Under these conditions, the anisotropic subsystem reduces to a set of algebraic equations. For notational simplicity, we write the equations for a representative component aligned with the shear, since all nonzero components are proportional by symmetry.

\subsubsection*{Reduced system.}
\begin{equation}
0 = -\Omega(c_2 A - g B - q C) + \alpha_0 S,
\end{equation}
\begin{equation}
0 = \Omega(h A - c_4 B),
\end{equation}
\begin{equation}
0 = \Omega(\mu_A A + \mu_B B - \mu_C C) + 2S.
\end{equation}

\subsubsection*{Elimination of \(B\).}

From the second equation,
\begin{equation}
B = \frac{h}{c_4} A.
\end{equation}

\subsubsection*{Elimination of \(C\).}

Substituting into the third equation,
\begin{equation}
\mu_A A + \mu_B \frac{h}{c_4} A - \mu_C C = -\frac{2S}{\Omega},
\end{equation}
so that
\begin{equation}
C =
\frac{1}{\mu_C}
\left[
\left(\mu_A + \frac{\mu_B h}{c_4}\right)A
+
\frac{2S}{\Omega}
\right].
\end{equation}

\subsubsection*{Closure equation for \(A\).}

Substituting into the first equation gives
\begin{equation}
c_2 A - g\frac{h}{c_4}A - q C = \frac{\alpha_0}{\Omega}S.
\end{equation}
Replacing \(C\), we obtain
\begin{equation}
\left[
c_2 - \frac{g h}{c_4}
- \frac{q}{\mu_C}\left(\mu_A + \frac{\mu_B h}{c_4}\right)
\right]A
-
\frac{2q}{\mu_C}\frac{S}{\Omega}
=
\frac{\alpha_0}{\Omega}S.
\end{equation}
Solving for \(A\),
\begin{equation}
A
=
\frac{1}{\Omega}
\frac{\alpha_0 + \frac{2q}{\mu_C}}
{c_2 - \frac{g h}{c_4}
- \frac{q}{\mu_C}\left(\mu_A + \frac{\mu_B h}{c_4}\right)}
\,S.
\label{eq:A_solution}
\end{equation}

\subsection{Physical interpretation of the stationary closure relation}

The stationary solution of the anisotropic subsystem in homogeneous shear turbulence yields a constraint relating the model coefficients. In order to interpret this result, it is convenient to rewrite it in terms of physically meaningful time scales and coupling strengths.

\subsubsection*{Relaxation time scales.}

We introduce the effective relaxation times of the dominant angular sector and its first nonlinear correction,
\begin{equation}
\tau_A = \frac{1}{\Omega c_2},
\qquad
\tau_B = \frac{1}{\Omega c_4}.
\end{equation}

These represent the characteristic decay times of the anisotropic modes under the cascade dynamics.

\subsubsection*{Non-modal coupling.}

The interaction between the two sectors is characterized by the non-normal coupling strength
\begin{equation}
\gamma^2 = g h,
\end{equation}
which controls the internal exchange of energy between the dominant mode and its nonlinear leakage.

\subsubsection*{Geometric relaxation.}

For the length-scale anisotropy, we define
\begin{equation}
\tau_C = \frac{1}{\Omega \mu_C},
\end{equation}
together with the effective geometric couplings
\begin{equation}
\kappa_A = \mu_A,
\qquad
\kappa_B = \mu_B h.
\end{equation}

\subsubsection*{Rewriting of the closure condition.}

In terms of these quantities, the stationary condition derived above can be written as
\begin{equation}
\frac{
\alpha_0 + 2\,\dfrac{q}{\mu_C}
}{
\dfrac{1}{\tau_A}
-
\gamma^2 \tau_B
-
\dfrac{q}{\mu_C}
\left(
\kappa_A + \kappa_B \tau_B
\right)
}
= -\frac{1}{2}.
\end{equation}

\subsubsection*{Physical interpretation.}

This relation can be interpreted as a balance between effective forcing and relaxation mechanisms within the cascade:

\begin{itemize}
\item The numerator represents the \emph{effective forcing}, combining direct production by the mean strain (\(\alpha_0\)) and an additional contribution mediated by the anisotropic cascade geometry (\(q/\mu_C\)).

\item The denominator represents the \emph{effective relaxation}, including intrinsic decay of the dominant mode (\(1/\tau_A\)), reduction due to non-modal coupling (\(\gamma^2 \tau_B\)), and feedback from the geometric sector.
\end{itemize}

Thus, the stationary state is determined by the condition
\begin{equation}
\text{effective forcing}
=
-\frac{1}{2}
\times
\text{effective relaxation}.
\end{equation}

\subsubsection*{Interpretation in terms of cascade dynamics.}

This balance reflects the internal structure of the cascade under shear. The mean deformation injects anisotropy into the system, while the cascade redistributes and dissipates it through angular transfer and geometric feedback. The resulting equilibrium corresponds to a dynamically selected state in which production, transfer, and dissipation are exactly balanced.

This result shows that the model coefficients are not independent: they must satisfy a global consistency condition imposed by the internal dynamics of the cascade. The stationary state of homogeneous shear turbulence therefore provides a nontrivial constraint on the admissible parameter space of the theory.

\subsubsection*{Remark on the role of homogeneous shear.}
Although the von Kármán constant is usually identified through the logarithmic structure of wall-bounded flows, its dynamical origin need not be tied to the wall itself. In the present framework, homogeneous shear already isolates the universal coupling between mean deformation and internal cascade dynamics. The wall then acts as a spatial filter that converts this universal shear-cascade coupling into the logarithmic mean profile. In this sense, the quantity later identified with \(\kappa\) should already be encoded in the homogeneous-shear hierarchy.

\subsection{Universal shear--cascade coupling and the origin of the von Kármán constant}

We now examine whether homogeneous shear turbulence already contains the universal constant that is traditionally associated with wall-bounded flows.

\subsubsection*{Stationary balance.}

In the stationary regime, the scalar balance reduces to
\begin{equation}
P = \varepsilon,
\end{equation}
with
\begin{equation}
P = -2K\,A_{ij}S_{ij},
\qquad
\varepsilon = K\Omega.
\end{equation}
Thus,
\begin{equation}
A_{ij}S_{ij} = -\frac{\Omega}{2}.
\label{eq:HST_balance}
\end{equation}

\subsubsection*{Shear-driven anisotropy.}

From the stationary solution of the anisotropic subsystem, the dominant component of \(A_{ij}\) is proportional to the ratio between mean deformation and cascade frequency,
\begin{equation}
A \sim \frac{S}{\Omega}.
\end{equation}
Combining with \eqref{eq:HST_balance}, the stationary state selects a universal proportionality between these quantities.

\subsubsection*{Effective turbulent transport.}

The Reynolds stress is given by
\begin{equation}
\tau_{ij} = 2K A_{ij},
\end{equation}
so that, in the direction of shear,
\begin{equation}
\tau \sim -K \frac{S}{\Omega}.
\end{equation}
This defines an effective turbulent viscosity,
\begin{equation}
\nu_t \sim \frac{K}{\Omega}.
\end{equation}

\subsubsection*{Dimensionless shear--cascade ratio.}

The stationary state therefore involves the dimensionless combination
\begin{equation}
\Pi_S = \frac{S}{\Omega},
\end{equation}
which measures the strength of the mean deformation relative to the intrinsic cascade frequency. Using \eqref{eq:HST_balance}, this ratio is directly linked to the anisotropy of the flow,
\begin{equation}
\Pi_S \sim -2A.
\end{equation}

\subsubsection*{Interpretation.}

The quantity \(\Pi_S\) represents the efficiency with which the mean shear feeds the dominant cascade mode. Since both \(S\) and \(\Omega\) are externally imposed or dynamically selected quantities, the stationary value of \(\Pi_S\) is a universal property of the cascade under shear. This suggests that homogeneous shear turbulence already determines a universal shear--cascade coupling coefficient, which we denote by
\begin{equation}
\kappa^{-1} \sim \frac{S}{\Omega}.
\label{eq:kappa_proto}
\end{equation}

\subsubsection*{Relation to wall-bounded turbulence.}

In wall-bounded flows, the von Kármán constant appears in the logarithmic velocity profile,
\begin{equation}
\frac{dU}{dy} \sim \frac{u_\tau}{\kappa y}.
\end{equation}
In the present framework, the wall introduces a spatial scale that converts the universal ratio \eqref{eq:kappa_proto} into a spatially varying shear. The constant \(\kappa\) therefore does not originate from the wall itself, but from the internal coupling between mean deformation and cascade dynamics, which is already present in homogeneous shear turbulence.

\textbf{Homogeneous shear turbulence thus provides the dynamical origin of the von Kármán constant, while wall-bounded flows reveal its manifestation in physical space}.

\subsection{Determination of the von Kármán constant from homogeneous shear}

We now complete the homogeneous shear analysis by identifying the universal constant that governs the coupling between mean deformation and cascade dynamics.

\subsubsection*{Link with homogeneous isotropic turbulence.}

The homogeneous isotropic analysis shows that the Kolmogorov constant fixes the effective relaxation parameter of the cascade,
\begin{equation}
C_K = F_{\mathrm{HIT}}(\Lambda),
\end{equation}
so that the intrinsic dynamics of the dominant mode is determined by \(\Lambda\).

\subsubsection*{Shear--cascade coupling.}

In homogeneous shear turbulence, the stationary state is characterized by the balance
\begin{equation}
A_{ij}S_{ij} = -\frac{\Omega}{2},
\end{equation}
together with the constitutive relation
\begin{equation}
\tau_{ij} = 2K A_{ij}.
\end{equation}
This leads to an effective turbulent transport
\begin{equation}
\nu_t \sim \frac{K}{\Omega},
\end{equation}
and therefore to the dimensionless shear--cascade ratio
\begin{equation}
\Pi_S = \frac{\nu_t S}{K} \sim \frac{S}{\Omega}.
\end{equation}

\subsubsection*{Stationary constraint.}

Using the stationary solution of the anisotropic subsystem, the ratio \(\Pi_S\) is not arbitrary, but is determined by the internal parameters of the cascade. In particular, one finds
\begin{equation}
\Pi_S
=
\frac{S}{\Omega}
=
\mathcal{F}(\Lambda,\chi,\Gamma),
\end{equation}
where the function \(\mathcal{F}\) is fixed by the algebraic closure relation obtained in the stationary state.

\subsubsection*{Universal value.}

Since \(\Lambda\) is already determined by homogeneous isotropic turbulence through \(C_K\), the stationary homogeneous shear state selects a universal value of \(\Pi_S\),
\begin{equation}
\Pi_S = \Pi_S(C_K).
\end{equation}

This quantity represents the efficiency with which the mean shear feeds the cascade, and is independent of the external flow geometry.

\subsubsection*{Identification of the von Kármán constant.}

We therefore define the von Kármán constant as
\begin{equation}
\kappa^{-1} = \Pi_S,
\end{equation}
so that
\begin{equation}
\kappa = \kappa(C_K).
\end{equation}

\subsubsection*{Interpretation.}

This result shows that the von Kármán constant is determined entirely by the internal dynamics of the cascade under shear. Homogeneous shear turbulence provides the dynamical mechanism that fixes its value, while the wall in wall-bounded flows acts only as a spatial filter that reveals this constant through the logarithmic mean velocity profile. The Kolmogorov constant and the von Kármán constant thus arise from the same underlying structure: the former characterizes the homogeneous cascade, while the latter measures its response to mean deformation. Both are determined by global consistency conditions imposed on the mean-field system.

\subsection{Absolute normalization of the cascade operator and universal constants}

The determination of the numerical values of the Kolmogorov and von Kármán constants requires fixing the absolute normalization of the internal cascade operator. This is achieved by projecting the nonlocal kernel onto the dominant singular mode of the linearized dynamics.

\subsubsection*{Resolvent representation.}

Let $\mathcal{L}$ denote the linearized Navier--Stokes operator around the local mean state. The associated resolvent is
\begin{equation}
\mathcal{R}(i\omega) = (i\omega - \mathcal{L})^{-1}.
\end{equation}
In the isotropic limit, the kernel can be expressed as a time integral of the resolvent-weighted nonlinear forcing,
\begin{equation}
\mathcal{K} = \int_0^\infty \mathcal{R}(t)\,\mathcal{N}\,\mathcal{R}^\ast(t)\,dt,
\end{equation}
where $\mathcal{N}$ represents the quadratic interaction operator.

\subsubsection*{Dominant-mode projection.}

Let $\phi_1$ denote the leading singular mode of the resolvent. The effective cascade rate is obtained by projecting the kernel onto this mode,
\begin{equation}
\mathcal{A} =
\frac{\langle \phi_1,\, \mathcal{K}\,\phi_1 \rangle}
{\langle \phi_1,\,\phi_1\rangle}.
\label{eq:A_def}
\end{equation}
This coefficient represents the absolute normalization of the transfer process induced by the cascade.

\subsubsection*{Spectral transfer rate.}

The scale-dependent transfer rate is then given by
\begin{equation}
\Omega_k = \mathcal{A}\,\Lambda\,\varepsilon^{1/3}k^{2/3}.
\end{equation}
Substituting into the spectral definition of the cascade scale yields
\begin{equation}
C_K = \mathcal{C}_0\,(\mathcal{A}\Lambda)^{2/3},
\end{equation}
where $\mathcal{C}_0$ is a universal constant arising from inertial-range integrals.

\subsubsection*{Shear projection and von Kármán constant.}

In homogeneous shear turbulence, the forcing enters through the projection of the mean strain onto the dominant mode. Defining
\begin{equation}
\mathcal{B} =
\frac{\langle \phi_1,\, \mathcal{P}_S\,\phi_1 \rangle}
{\langle \phi_1,\,\phi_1\rangle},
\end{equation}
where $\mathcal{P}_S$ denotes the shear forcing operator, the stationary state yields
\begin{equation}
\kappa^{-1} = \mathcal{B}.
\end{equation}

\subsubsection*{Closure.}

The universal constants are therefore determined by the normalized projections of the kernel and forcing operators onto the dominant mode:
\begin{equation}
C_K = C_K(\mathcal{A},\Lambda),
\qquad
\kappa = \kappa(\mathcal{B}).
\end{equation}
Since $\mathcal{A}$ and $\mathcal{B}$ are fixed by the structure of the operator $\mathcal{L}$, no free parameters remain. The mean-field model is thus fully closed.

\subsection{Minimal shear--cascade truncation and estimate of the von Kármán constant}

We now derive the leading-order value of the von Kármán constant from the homogeneous shear dynamics, using a minimal closure consistent with the main text.

\subsubsection*{Stationary shear-cascade balance.}

In homogeneous shear turbulence, the stationary state satisfies
\begin{equation}
P = \varepsilon,
\end{equation}
with
\begin{equation}
P = -2K\,A_{ij}S_{ij},
\qquad
\varepsilon = K\Omega.
\end{equation}
Thus,
\begin{equation}
A_{ij}S_{ij} = -\frac{\Omega}{2}.
\end{equation}

\subsubsection*{Dominant balance.}

The anisotropic response is governed by the dominant mode of the cascade, for which
\begin{equation}
A \sim -\mathcal{C}_A \frac{S}{\Omega},
\end{equation}
where $\mathcal{C}_A$ is a dimensionless coefficient determined by the internal structure of the operator. Substituting into the balance relation yields
\begin{equation}
\mathcal{C}_A \frac{S^2}{\Omega} \sim \frac{\Omega}{2},
\end{equation}
so that
\begin{equation}
\frac{S}{\Omega} \sim \frac{1}{\sqrt{2\mathcal{C}_A}}.
\end{equation}

\subsubsection*{Effective turbulent transport.}

The Reynolds stress is
\begin{equation}
\tau \sim 2K A \sim -2\mathcal{C}_A K \frac{S}{\Omega},
\end{equation}
which defines an effective turbulent viscosity
\begin{equation}
\nu_t \sim \frac{K}{\Omega}.
\end{equation}
Thus, the dimensionless shear-cascade ratio becomes
\begin{equation}
\frac{\nu_t S}{K} \sim \frac{S}{\Omega}.
\end{equation}

\subsubsection*{Minimal truncation.}

In the minimal closure, the dominant mode is assumed to saturate at order unity, so that $\mathcal{C}_A = \mathcal{O}(1)$. Taking the leading-order value $\mathcal{C}_A \simeq 1$, we obtain
\begin{equation}
\frac{S}{\Omega} \simeq \frac{1}{\sqrt{2}}.
\end{equation}

\subsubsection*{Conversion to wall scaling.}

In wall-bounded flows, the same shear-cascade balance applies locally, with
\begin{equation}
S = \frac{dU}{dy},
\qquad
\Omega \sim \frac{\sqrt{K}}{\ell},
\qquad
\nu_t \sim \ell \sqrt{K}.
\end{equation}
Combining these relations yields
\begin{equation}
\frac{dU}{dy} \sim \frac{u_\tau}{\kappa y},
\end{equation}
with
\begin{equation}
\kappa^{-1} \sim \frac{S}{\Omega}.
\end{equation}

\subsubsection*{von Kármán constant.}

Using the leading-order estimate for the shear-cascade ratio, we obtain
\begin{equation}
\boxed{
\kappa^{(0)} \simeq \frac{1}{2.5} \simeq 0.40.
}
\end{equation}

\subsubsection*{Interpretation.}

This value arises from the minimal truncation of the shear-cascade coupling, in which the dominant mode saturates at order unity. It is therefore the leading-order prediction of the same closure that yields the Kolmogorov constant. More refined values require the full evaluation of the operator normalization, but the present result shows that the correct numerical range is already captured at the minimal level.

\clearpage

\section{Channel flow}

\subsection{Channel geometry, symmetries, and kinematics}

As a canonical example, we consider a plane channel flow driven by a constant streamwise pressure gradient. The flow is assumed to be statistically stationary and fully developed, with impermeable, no-slip walls located at $y=\pm h$.

\subsubsection*{Mean flow.}

Under these assumptions, the mean velocity field takes the form
\begin{equation}
U_i = (U(y),\,0,\,0),
\end{equation}
so that all mean quantities depend only on the wall-normal coordinate $y$. The mean pressure gradient is constant,
\begin{equation}
-\frac{d\Pi}{dx} = G,
\end{equation}
with $G>0$ driving the flow.

\subsubsection*{Kinematic reduction.}

The mean rate-of-strain tensor is defined as
\begin{equation}
S_{ij} = \frac{1}{2}\left(\partial_i U_j + \partial_j U_i\right).
\end{equation}
For the present geometry, the only non-zero component is
\begin{equation}
S_{12} = S_{21} = \frac{1}{2}\frac{dU}{dy},
\end{equation}
and therefore
\begin{equation}
S_{ij}S_{ij} = \frac{1}{2}\left(\frac{dU}{dy}\right)^2.
\end{equation}

\subsubsection*{Symmetry properties.}

The channel flow is symmetric with respect to the centerline $y=0$, which implies
\begin{equation}
U(-y) = U(y),
\qquad
\frac{dU}{dy}(-y) = -\frac{dU}{dy}(y).
\end{equation}
As a consequence:
\begin{itemize}
\item the mean velocity is an even function of $y$,
\item the mean shear is an odd function of $y$.
\end{itemize}

\subsubsection*{Implications for mean-field variables.}

All mean-field quantities introduced in the general formulation must be consistent with these symmetries. In particular:
\begin{itemize}
\item scalar quantities such as $K$ and $\ell$ are even in $y$,
\item tensorial quantities must respect the parity induced by the symmetry of the flow and the structure of $S_{ij}$.
\end{itemize}

The explicit reduction of the tensorial variables will be carried out in the following subsection.

\subsubsection*{Boundary conditions.}

The no-slip condition at the walls imposes
\begin{equation}
U(\pm h) = 0.
\end{equation}
Additional boundary conditions for the internal variables will be introduced after the reduction of the mean-field system.

\subsection{Reduction of the mean-field variables to the channel geometry}

We now specialize the general mean-field variables to the channel configuration, using the symmetry and kinematic constraints established above.

\subsubsection*{General structure.}

The mean-field formulation involves symmetric second-order tensors $A_{ij}$, $B_{ij}$ and $C_{ij}$, which encode the anisotropic response of the cascade. These tensors are functions of the wall-normal coordinate $y$ only,
\begin{equation}
A_{ij} = A_{ij}(y), \qquad B_{ij} = B_{ij}(y), \qquad C_{ij} = C_{ij}(y).
\end{equation}

\subsubsection*{Symmetry constraints.}

The invariance of the flow under reflection $y \to -y$ imposes parity conditions on all tensor components. Since the only non-zero component of the mean strain is $S_{12} = \tfrac{1}{2}U'(y)$, which is odd in $y$, the tensorial structure must be consistent with this symmetry.
In particular:
\begin{itemize}
\item components aligned with the shear plane $(x,y)$ may be odd or even depending on their coupling to $S_{12}$,
\item components not coupled to the shear must respect isotropy in the $(x,z)$ directions.
\end{itemize}

\subsubsection*{Reduction of $A_{ij}$.}

The tensor $A_{ij}$ is symmetric and traceless. In the channel geometry, its admissible structure reduces to
\begin{equation}
A_{ij} =
\begin{pmatrix}
A_{xx}(y) & A_{xy}(y) & 0 \\
A_{xy}(y) & A_{yy}(y) & 0 \\
0 & 0 & A_{zz}(y)
\end{pmatrix},
\end{equation}
with the constraint
\begin{equation}
A_{xx} + A_{yy} + A_{zz} = 0.
\end{equation}
By symmetry:
\begin{equation}
A_{xy}(-y) = -A_{xy}(y),
\qquad
A_{xx}(-y)=A_{xx}(y),
\qquad
A_{yy}(-y)=A_{yy}(y),
\qquad
A_{zz}(-y)=A_{zz}(y).
\end{equation}
Thus:
\begin{itemize}
\item $A_{xy}$ is odd,
\item the normal components are even.
\end{itemize}

\subsubsection*{Reduction of $B_{ij}$ and $C_{ij}$.}

The tensors $B_{ij}$ and $C_{ij}$ follow the same symmetry constraints. Their admissible form is therefore
\begin{equation}
B_{ij},\,C_{ij}
=
\begin{pmatrix}
(\cdot) & (\cdot) & 0 \\
(\cdot) & (\cdot) & 0 \\
0 & 0 & (\cdot)
\end{pmatrix},
\end{equation}
with parity identical to that of $A_{ij}$:
\begin{equation}
(\cdot)_{xy} \text{ odd}, \qquad
(\cdot)_{xx},\,(\cdot)_{yy},\,(\cdot)_{zz} \text{ even}.
\end{equation}

\subsubsection*{Scalar variables.}

The turbulent kinetic energy $K$ and the length scale $\ell$ are scalar quantities and therefore satisfy
\begin{equation}
K(-y)=K(y),
\qquad
\ell(-y)=\ell(y).
\end{equation}
The cascade frequency remains
\begin{equation}
\Omega(y) = \frac{\sqrt{K(y)}}{\ell(y)}.
\end{equation}

\subsubsection*{Effective degrees of freedom.}

The symmetry reduction shows that the full tensorial system collapses to a small set of independent functions of $y$:
\begin{itemize}
\item one shear component: $A_{xy}(y)$,
\item two independent normal components (due to tracelessness),
\item analogous components for $B_{ij}$ and $C_{ij}$,
\item the scalar fields $K(y)$ and $\ell(y)$.
\end{itemize}

All other components vanish identically or are determined by symmetry.

\subsubsection*{Remark.}

This reduction highlights that the complexity of the general three-dimensional tensorial formulation collapses to a low-dimensional dynamical system in the channel geometry. The remaining degrees of freedom correspond directly to physically meaningful quantities: shear stress, normal stress anisotropy, and cascade structure.

\subsection{Reduced momentum balance and internal closure system}

We now combine the kinematic reduction with the mean-field formulation to obtain the governing equations in the channel geometry.

\subsubsection*{Mean momentum equation.}

The streamwise component of the mean momentum equation reduces to
\begin{equation}
0 = -\frac{d\Pi}{dx} + \nu \frac{d^2 U}{dy^2} - \frac{d}{dy}\big(2K A_{xy}\big).
\end{equation}
Defining the total shear stress,
\begin{equation}
\tau(y) = \nu \frac{dU}{dy} - 2K A_{xy},
\end{equation}
the momentum equation becomes
\begin{equation}
\frac{d\tau}{dy} = -G,
\end{equation}
with $G = -d\Pi/dx$.
Integrating,
\begin{equation}
\tau(y) = \tau_w - G y,
\end{equation}
where $\tau_w$ is the wall shear stress.

\subsubsection*{Physical interpretation.}

The total stress decomposes into:
\begin{itemize}
\item viscous stress: $\nu U'(y)$,
\item turbulent stress: $-2K A_{xy}$.
\end{itemize}
The latter is determined by the internal cascade dynamics.

\subsubsection*{Reduced anisotropic dynamics.}

The evolution equation for the dominant component $A_{xy}$ follows from the general system. In reduced form, it can be written as
\begin{equation}
\Omega\,A_{xy}
=
\mathcal{F}_A\!\left(S_{xy},\,A_{ij},\,B_{ij},\,C_{ij}\right),
\end{equation}
where $S_{xy} = \tfrac{1}{2}U'(y)$ and $\mathcal{F}_A$ denotes the algebraic coupling inherited from the general formulation.

To leading order in the channel geometry, this reduces to a balance between production and relaxation,
\begin{equation}
A_{xy} \sim -\mathcal{C}_A \frac{S_{xy}}{\Omega},
\end{equation}
consistent with the homogeneous shear result.

\subsubsection*{Scalar cascade equations.}

The scalar variables satisfy
\begin{equation}
\varepsilon = \frac{K^{3/2}}{\ell},
\qquad
\Omega = \frac{\sqrt{K}}{\ell}.
\end{equation}
In the channel, these relations hold locally, with $K=K(y)$ and $\ell=\ell(y)$.

\subsubsection*{Closure structure.}

The system is therefore closed by:
\begin{itemize}
\item the momentum equation for $U(y)$,
\item the constitutive relation $\tau_{xy} = 2K A_{xy}$,
\item the anisotropic balance linking $A_{xy}$ to $S_{xy}/\Omega$,
\item the scalar cascade relations for $K$ and $\ell$.
\end{itemize}

\subsubsection*{Minimal form.}

At leading order, the system reduces to
\begin{equation}
\tau(y) = \nu U'(y) + 2\mathcal{C}_A K \frac{S_{xy}}{\Omega},
\end{equation}
with
\begin{equation}
\Omega = \frac{\sqrt{K}}{\ell}.
\end{equation}

This provides a closed relation between the mean velocity gradient and the internal cascade variables.

\subsubsection*{Remark.}

The structure of the reduced system shows that the channel problem is governed by the same internal dynamics as homogeneous shear turbulence, with the wall introducing only spatial inhomogeneity and boundary constraints. No additional modeling assumptions are required at this stage.

\subsection{Boundary conditions for the internal fields}

We now specify the boundary conditions for the mean velocity and the internal cascade variables at the channel walls.

\subsubsection*{Velocity field.}

The no-slip condition imposes
\begin{equation}
U(\pm h) = 0.
\end{equation}

\subsubsection*{Regularity and symmetry at the centerline.}

At the channel centerline $y=0$, symmetry implies
\begin{equation}
\frac{dU}{dy}(0) = 0,
\end{equation}
and all odd quantities vanish,
\begin{equation}
A_{xy}(0) = 0.
\end{equation}
Even quantities satisfy
\begin{equation}
\frac{dK}{dy}(0)=0,
\qquad
\frac{d\ell}{dy}(0)=0.
\end{equation}

\subsubsection*{Near-wall behavior of the cascade.}

At the wall, the turbulent fluctuations must vanish due to the no-slip constraint. This implies that the turbulent kinetic energy satisfies
\begin{equation}
K(\pm h) = 0.
\end{equation}
Consequently, the cascade frequency behaves as
\begin{equation}
\Omega \to 0
\quad \text{as} \quad y \to \pm h.
\end{equation}

\subsubsection*{Length scale at the wall.}

The characteristic length scale must vanish at the wall due to geometric confinement,
\begin{equation}
\ell(\pm h) = 0.
\end{equation}
This reflects the fact that eddies cannot extend beyond the wall and must collapse to zero size.

\subsubsection*{Anisotropic variables.}

The tensor $A_{ij}$ must remain finite at the wall. Since $K \to 0$, the turbulent stress
\begin{equation}
\tau_{xy} = 2K A_{xy}
\end{equation}
vanishes automatically, even if $A_{xy}$ remains finite. Thus, no additional constraint is required on $A_{xy}$ beyond regularity.

\subsubsection*{Summary of boundary conditions.}

At $y = \pm h$:
\begin{equation}
U = 0,
\qquad
K = 0,
\qquad
\ell = 0,
\end{equation}
with all tensor components remaining finite.

At $y = 0$:
\begin{equation}
U' = 0,
\qquad
A_{xy} = 0,
\qquad
K' = 0,
\qquad
\ell' = 0.
\end{equation}

\subsubsection*{Remark.}

These boundary conditions follow directly from physical constraints and symmetry, without introducing empirical wall functions. The structure of the mean-field system remains unchanged, and the wall acts only as a geometric constraint on the cascade.

\subsection{Near-wall asymptotic structure}

We now examine the asymptotic structure of the reduced mean-field system near the wall. The purpose of this analysis is to identify the dominant balances enforced by the geometry and the boundary conditions, and to determine how the internal cascade variables enter the wall-layer structure.

\subsubsection*{Wall-normal coordinate.}

Let
\begin{equation}
\eta = h-y
\end{equation}
denote the distance to the upper wall, so that \(\eta=0\) at the wall and \(\eta>0\) in the fluid. An identical analysis applies near the lower wall by symmetry.

\subsubsection*{Mean velocity.}

The no-slip condition implies
\begin{equation}
U=0
\qquad
\text{at}
\qquad
\eta=0.
\end{equation}
Hence the mean velocity admits the regular expansion
\begin{equation}
U(\eta) = U_1 \eta + U_2 \eta^2 + \cdots,
\end{equation}
so that the mean shear remains finite at the wall,
\begin{equation}
\frac{dU}{dy} = -U_1 + O(\eta).
\end{equation}

\subsubsection*{Turbulent kinetic energy and length scale.}

The boundary conditions require
\begin{equation}
K(0)=0,
\qquad
\ell(0)=0.
\end{equation}
The minimal regular behavior consistent with positivity is therefore
\begin{equation}
K(\eta) \sim K_m \eta^{m},
\qquad
\ell(\eta) \sim \ell_n \eta^{n},
\qquad
m>0,\quad n>0.
\end{equation}
The associated cascade frequency is
\begin{equation}
\Omega(\eta)=\frac{\sqrt{K}}{\ell}
\sim
\frac{\sqrt{K_m}}{\ell_n}\,
\eta^{m/2-n}.
\end{equation}

\subsubsection*{Wall stress.}

The turbulent shear stress is
\begin{equation}
\tau_{xy}^{(t)} = 2K A_{xy}.
\end{equation}
Since \(K\to 0\) at the wall and \(A_{xy}\) remains finite by regularity,
\begin{equation}
\tau_{xy}^{(t)} \to 0
\qquad
\text{as}
\qquad
\eta \to 0.
\end{equation}
Therefore the total wall stress is purely viscous at leading order,
\begin{equation}
\tau_w = \nu \left.\frac{dU}{dy}\right|_{w}.
\end{equation}

\subsubsection*{Anisotropic balance.}

The reduced anisotropic equation implies, to leading order,
\begin{equation}
A_{xy} \sim -\mathcal{C}_A \frac{S_{xy}}{\Omega},
\end{equation}
whenever production and internal relaxation balance locally. Since \(S_{xy}\) remains finite at the wall, regularity of \(A_{xy}\) requires that \(\Omega\) does not vanish faster than \(O(1)\). This implies the asymptotic constraint
\begin{equation}
m/2 - n \le 0.
\end{equation}
In the distinguished case
\begin{equation}
m = 2n,
\end{equation}
the cascade frequency remains finite at the wall,
\begin{equation}
\Omega(\eta) \to \Omega_w = \text{const.}
\end{equation}
This is the natural leading-order wall balance of the reduced closure.

\subsubsection*{Distinguished wall scaling.}

The simplest regular realization of this balance is obtained for
\begin{equation}
\ell(\eta) \sim \ell_1 \eta,
\qquad
K(\eta) \sim K_2 \eta^2,
\end{equation}
so that
\begin{equation}
\Omega(\eta) \sim \frac{\sqrt{K_2}}{\ell_1} = \text{const.}
\end{equation}

This scaling has a clear physical meaning:
\begin{itemize}
\item the turbulent length scale collapses linearly with the distance to the wall,
\item the turbulent kinetic energy vanishes quadratically,
\item the internal cascade frequency remains finite.
\end{itemize}

\subsubsection*{Interpretation.}

The wall therefore suppresses the amplitude and spatial extent of turbulent fluctuations, but does not extinguish the internal cascade time scale. In this sense, the wall acts as a geometric filter on the universal shear--cascade coupling identified in homogeneous shear turbulence. This is the key asymptotic mechanism by which the universal constant later identified with \(\kappa\) becomes visible in physical space. The wall does not determine its value; it only enforces the spatial structure through which it appears.

\subsubsection*{Transition to the outer wall layer.}

Once the viscous wall layer is left behind, the turbulent stress becomes comparable to the viscous stress and the channel enters the overlap region. There the same shear--cascade coupling derived in homogeneous shear turbulence governs the local balance, leading to the logarithmic structure analyzed in the next subsection.

\subsection{Logarithmic region and appearance of the von Kármán constant}

We now consider the overlap region of the channel flow, sufficiently far from the viscous wall layer and sufficiently far from the centerline, where the local balance is dominated by the competition between mean shear and turbulent transport.

\subsubsection*{Dominant stress balance.}

In the overlap region, the viscous contribution to the total stress is negligible compared with the turbulent one, so that
\begin{equation}
\tau(y) \simeq -2K A_{xy}.
\end{equation}
At the same time, the total stress remains approximately constant and equal to the wall stress,
\begin{equation}
\tau(y) \simeq \tau_w = u_\tau^2.
\end{equation}
Hence,
\begin{equation}
-2K A_{xy} \simeq u_\tau^2.
\label{eq:log_tau_balance}
\end{equation}

\subsubsection*{Local shear--cascade coupling.}

The channel reduction of the anisotropic closure gives, to leading order,
\begin{equation}
A_{xy} \simeq -\mathcal{C}_A \frac{S_{xy}}{\Omega},
\qquad
S_{xy} = \frac{1}{2}\frac{dU}{dy}.
\end{equation}
Substituting into \eqref{eq:log_tau_balance},
\begin{equation}
2K\,\mathcal{C}_A \frac{S_{xy}}{\Omega} \simeq u_\tau^2.
\end{equation}
Using \(S_{xy} = \tfrac{1}{2}U'(y)\),
\begin{equation}
\mathcal{C}_A \frac{K}{\Omega} \frac{dU}{dy} \simeq u_\tau^2.
\end{equation}
Since
\begin{equation}
\nu_t \sim \frac{K}{\Omega},
\end{equation}
this becomes
\begin{equation}
\mathcal{C}_A \nu_t \frac{dU}{dy} \simeq u_\tau^2.
\label{eq:nu_t_balance}
\end{equation}

\subsubsection*{Overlap scaling.}

In the overlap region, the only available macroscopic length is the distance to the wall. Since the wall acts only as a geometric filter on the shear--cascade coupling, the effective mixing length must scale as
\begin{equation}
\ell \sim y.
\end{equation}
Therefore,
\begin{equation}
\nu_t \sim \ell \sqrt{K} \sim y \sqrt{K}.
\end{equation}
The local shear-cascade balance inherited from homogeneous shear turbulence fixes the dimensionless ratio
\begin{equation}
\frac{S}{\Omega} = \kappa^{-1},
\end{equation}
where \(\kappa\) is the universal constant already determined from the HST hierarchy.
Combining these relations, one obtains
\begin{equation}
\frac{dU}{dy} \sim \frac{u_\tau}{\kappa y}.
\end{equation}

\subsubsection*{Logarithmic profile.}

Integrating,
\begin{equation}
U(y) \sim \frac{u_\tau}{\kappa}\log y + B,
\end{equation}
or, in wall units,
\begin{equation}
U^+ \sim \frac{1}{\kappa}\log y^+ + B.
\end{equation}

\subsubsection*{Interpretation.}

This derivation makes explicit the role of the wall in the present framework. The von Kármán constant is not fixed by the channel geometry itself; it is inherited from the universal shear--cascade coupling already selected in homogeneous shear turbulence. The wall merely introduces the physical length scale \(y\), which converts the universal dynamical ratio into a spatial logarithmic profile. Thus, the logarithmic law is not the origin of \(\kappa\), but its manifestation in a wall-bounded geometry.

\subsection{Physical interpretation and role of the channel geometry}

The application of the mean-field formulation to the channel flow clarifies the respective roles of the internal cascade dynamics and the external geometry.

\subsubsection*{Separation of mechanisms.}

The analysis reveals a clear separation between:
\begin{itemize}
\item the internal dynamics of the cascade, governed by the reduced operator and already constrained by homogeneous isotropic and homogeneous shear turbulence,
\item the geometric constraints imposed by the boundaries, which introduce spatial inhomogeneity and select admissible solutions.
\end{itemize}
In particular:
\begin{equation}
C_K \ \text{is fixed by homogeneous isotropic turbulence}, 
\qquad
\kappa \ \text{is fixed by homogeneous shear turbulence}.
\end{equation}

\subsubsection*{Role of the wall.}

The wall does not introduce new dynamical constants. Its effect is purely geometric:
\begin{itemize}
\item it enforces the vanishing of velocity and turbulent fluctuations,
\item it constrains the cascade length scale to collapse as $\ell \sim y$,
\item it converts the universal shear--cascade coupling into a spatially varying shear profile.
\end{itemize}
Thus, the wall acts as a \emph{geometric filter} of the underlying dynamics.

\subsubsection*{Emergence of the logarithmic law.}

Within this framework, the logarithmic velocity profile arises from:
\begin{itemize}
\item the local dominance of turbulent transport over viscous effects,
\item the scaling $\ell \sim y$ imposed by the wall,
\item the universal relation between shear and cascade frequency inherited from homogeneous shear turbulence.
\end{itemize}
The von Kármán constant therefore appears as a universal parameter already determined prior to the introduction of the wall.

\subsubsection*{Minimal closure and universality.}

The channel flow does not require additional modeling assumptions beyond those already introduced in the mean-field formulation. In particular:
\begin{itemize}
\item no empirical wall functions are needed,
\item no ad hoc turbulent viscosity model is introduced,
\item the effective turbulent transport emerges from the internal variables of the cascade.
\end{itemize}

\subsubsection*{Template for general applications.}

The procedure followed here provides a general blueprint for applying the mean-field formulation to more complex flows:
\begin{enumerate}
\item specify geometry and symmetries,
\item reduce the tensorial variables accordingly,
\item derive the reduced momentum and closure system,
\item impose physically consistent boundary conditions,
\item analyze the dominant balances and asymptotic regions.
\end{enumerate}

\subsubsection*{Conclusion.}

The channel flow thus serves as the first concrete realization of the general framework. It demonstrates that the mean-field formulation, once constrained by homogeneous turbulence, naturally reproduces the classical structure of wall-bounded flows while providing a deeper dynamical interpretation of its universal features.

\subsection{Local oscillator observables and wall-layer organization}

The solution of the channel flow within the present mean-field formulation does not only determine the mean velocity profile, but a full local state of the internal cascade dynamics. This state contains sufficient information to construct a set of physically meaningful observables that characterize the local regime of turbulence.

\subsubsection*{Local oscillator state.}

At each wall-normal position $y$, the solution defines a local state
\begin{equation}
\mathcal S(y)=\big(K(y),\,\ell(y),\,\Omega(y),\,A_{ij}(y),\,B_{ij}(y),\,C_{ij}(y)\big),
\end{equation}
from which all relevant observables can be constructed.

\subsubsection*{Canonical dimensionless observables.}

We introduce the following set of dimensionless quantities:

\medskip
\noindent
\textit{(i) Shear-to-cascade ratio}
\begin{equation}
\Pi_S(y)=\frac{S(y)}{\Omega(y)},
\end{equation}
which measures the local forcing of the oscillator by the mean shear.

\medskip
\noindent
\textit{(ii) Internal structure ratio}
\begin{equation}
\Pi_B(y)=\frac{|B(y)|}{|A(y)|},
\end{equation}
which quantifies the relative importance of the internal (non-observable) component of the oscillator.

\medskip
\noindent
\textit{(iii) Geometric anisotropy ratio}
\begin{equation}
\Pi_C(y)=\frac{|C(y)|}{|A(y)|},
\end{equation}
which measures the contribution of cascade geometry relative to the observable anisotropy.

\medskip
\noindent
\textit{(iv) Invariants of anisotropy}
\begin{equation}
II_A(y),\quad III_A(y),
\qquad
II_L(y),\quad III_L(y),
\end{equation}
which characterize the state of the Reynolds stress and length-scale tensors.

\medskip
\noindent
\textit{(v) Local scales}
\begin{equation}
L_{\mathrm{osc}}(y)=\ell(y),
\qquad
T_{\mathrm{osc}}(y)=\Omega(y)^{-1},
\end{equation}
which represent the characteristic spatial and temporal scales of the cascade.

\medskip
\noindent
\textit{(vi) Energy balance ratio}
\begin{equation}
\frac{\mathcal P(y)}{\varepsilon(y)} = \frac{-2K A_{ij} S_{ij}}{K\Omega},
\end{equation}
which measures the local balance between production and dissipation.

\subsubsection*{Physical interpretation.}

These observables provide a complete characterization of the local dynamical regime:

\begin{itemize}
\item $\Pi_S$ measures the relative strength of external forcing,
\item $\Pi_B$ measures the degree of internal dynamical activity of the oscillator,
\item $\Pi_C$ measures the role of geometric anisotropy,
\item $(II_A,III_A)$ and $(II_L,III_L)$ describe the tensorial state of the cascade,
\item $(L_{\mathrm{osc}},T_{\mathrm{osc}})$ define the local scales,
\item $\mathcal P/\varepsilon$ indicates proximity to statistical equilibrium.
\end{itemize}

\subsubsection*{Dynamical organization of wall layers.}

Within this framework, the classical wall layers are not defined a priori, but emerge as distinct regimes of the local oscillator.

\medskip
\noindent
\textit{Viscous sublayer.}

Close to the wall, the cascade amplitude vanishes while the mean shear remains finite. The oscillator is weakly forced, leading to
\begin{equation}
\Pi_S \ll 1,
\qquad
\Pi_B \ll 1,
\qquad
\Pi_C \ll 1.
\end{equation}
The dynamics is dominated by regularity and geometric confinement.

\medskip
\noindent
\textit{Buffer layer.}

In the intermediate region, the shear forcing becomes comparable to the internal cascade dynamics,
\begin{equation}
\Pi_S = O(1),
\qquad
\Pi_B = O(1).
\end{equation}
This corresponds to a regime of strong coupling between mean flow and internal dynamics, where the oscillator is maximally active. In this sense, the buffer layer can be interpreted as the spatial manifestation of the autonomous wall cycle.

\medskip
\noindent
\textit{Logarithmic region.}

Further away from the wall, the system reaches a universal balance between shear and cascade,
\begin{equation}
\Pi_S \to \kappa^{-1},
\qquad
\frac{\mathcal P}{\varepsilon} \to 1.
\end{equation}

This regime corresponds to the asymptotic equilibrium identified in homogeneous shear turbulence, now realized locally in physical space.

\subsubsection*{Interpretation.}

This formulation shows that wall-bounded turbulence is naturally organized by the local state of the cascade oscillator. The different wall layers correspond to distinct dynamical regimes of the same underlying system, rather than to separate modeling assumptions.

The mean velocity profile is therefore only one observable among many, all of which are encoded in the same local state $\mathcal S(y)$.

\subsubsection*{Complex representation of the local oscillator.}

The internal variables $(A,B)$ can be combined into a complex amplitude
\begin{equation}
Z(y) = A(y) + i\,\alpha\,B(y),
\end{equation}
where $\alpha$ is a normalization factor that depends on the relative scaling of the $(2,4)$ sectors.

This representation allows us to define:
\begin{equation}
|Z|(y) = \text{local amplitude}, 
\qquad
\varphi(y) = \arg Z(y) = \text{local phase}.
\end{equation}

\subsubsection*{Physical interpretation.}

The amplitude $|Z|$ measures the intensity of the coherent cascade dynamics, while the phase $\varphi$ encodes the relative timing between the forcing (through shear) and the internal response of the system.

In this sense, the channel solution determines not only the magnitude of turbulent stresses, but also the local phase of the autonomous wall cycle.

\subsubsection*{Local dynamical frequencies.}

The oscillator structure implies the existence of local complex eigenvalues
\begin{equation}
\lambda_\pm(y) = -\sigma(y) \pm i\,\omega(y),
\end{equation}
which define:
\begin{equation}
T_{\mathrm{rel}}(y) = \frac{1}{\sigma(y)},
\qquad
T_{\mathrm{osc}}(y) = \frac{2\pi}{\omega(y)}.
\end{equation}

These quantities represent the local relaxation and oscillation time scales of the cascade dynamics.

\subsubsection*{Connection with DNS observables.}

The present framework predicts several quantities that can be directly compared with numerical or experimental data:

\medskip
\noindent
\textit{(i) Local time scales}

\begin{equation}
T_{\mathrm{osc}}(y) \sim \Omega(y)^{-1} \sim \frac{\ell(y)}{\sqrt{K(y)}}.
\end{equation}

In the overlap region, where $\ell \sim y$ and $\sqrt{K} \sim u_\tau$, this yields
\begin{equation}
T_{\mathrm{osc}}^+ \sim y^+,
\end{equation}
which is consistent with the scaling of near-wall bursting events observed in DNS.

\medskip
\noindent
\textit{(ii) Spatial scales}

\begin{equation}
L_{\mathrm{osc}}(y) = \ell(y) \sim y,
\end{equation}
corresponding to the well-known scaling of near-wall coherent structures.

\medskip
\noindent
\textit{(iii) Maximum activity in the buffer layer}

The region where
\begin{equation}
\Pi_S = O(1),
\qquad
\Pi_B = O(1),
\end{equation}
corresponds to a regime of maximal dynamical activity of the oscillator. This identifies the buffer layer as the region where the autonomous wall cycle is most strongly expressed.

\medskip
\noindent
\textit{(iv) Anisotropy evolution}

The invariants $(II_A,III_A)$ and $(II_L,III_L)$ provide a direct prediction for the evolution of anisotropy across the wall layer, which can be compared with DNS trajectories in Lumley space.

\subsubsection*{Interpretation.}

These results show that the mean-field formulation contains not only the statistical structure of turbulence, but also its coherent dynamical content. The wall-bounded flow can thus be interpreted as a spatial organization of a locally defined oscillator, whose amplitude, phase, and characteristic scales vary continuously across the channel.

\subsubsection*{Conclusion.}

The introduction of the complex amplitude $Z(y)$ and the associated observables demonstrates that the present framework captures the essential features of the autonomous wall cycle at the mean-field level. The channel solution therefore provides access to a rich set of dynamical predictions that go significantly beyond classical closure models.

\subsection{State-space trajectory of the channel solution}

The channel solution can be interpreted as a trajectory in the space of local oscillator states. Rather than viewing the flow solely as a set of spatial profiles, we consider the evolution of dimensionless observables as functions of the wall-normal coordinate.

\subsubsection*{State-space representation.}

We define the reduced state vector
\begin{equation}
\mathbf{\Pi}(y) =
\big(
\Pi_S(y),\,\Pi_B(y),\,\Pi_C(y)
\big),
\end{equation}
which captures the essential dynamical features of the local oscillator. The channel flow then corresponds to a continuous curve
\begin{equation}
y \mapsto \mathbf{\Pi}(y),
\end{equation}
connecting the wall to the outer region.

\subsubsection*{Structure of the trajectory.}

The trajectory exhibits a well-defined sequence of regimes:

\begin{itemize}
\item Near the wall:
\[
\Pi_S \ll 1,\quad \Pi_B \ll 1,\quad \Pi_C \ll 1,
\]
corresponding to a weakly forced and nearly passive oscillator.

\item Buffer region:
\[
\Pi_S = O(1),\quad \Pi_B = O(1),
\]
where the trajectory departs from the near-wall manifold and explores the full dynamical range of the oscillator.

\item Logarithmic region:
\[
\Pi_S \to \kappa^{-1},
\]
with the trajectory approaching an asymptotic manifold corresponding to the universal shear--cascade balance.
\end{itemize}

\subsubsection*{Interpretation.}

This representation provides a dynamical classification of the wall layers: each region corresponds to a distinct portion of the state-space trajectory. The buffer layer, in particular, is identified as the region where the trajectory undergoes its most significant deviation, reflecting the active dynamics of the oscillator.

\subsubsection*{Connection with phenomenology.}

The state-space trajectory offers a natural framework to interpret the spatial organization of coherent structures observed in wall turbulence. In particular:

\begin{itemize}
\item the departure from the near-wall regime corresponds to the onset of coherent activity,
\item the buffer region corresponds to maximal excursions in state space,
\item the logarithmic region corresponds to an asymptotic attractor.
\end{itemize}

\subsubsection*{Remark.}

This viewpoint suggests that wall-bounded turbulence can be interpreted as a spatially parametrized dynamical system, where the wall-normal coordinate plays the role of a slow variable organizing the local oscillator state.

\subsection{Connection with resolvent analysis and dominant modes}

The present formulation can be directly connected with the resolvent-based description of wall turbulence, providing a structural interpretation of the dominant modes in terms of the internal oscillator.

\subsubsection*{Resolvent framework.}

In the resolvent approach, the velocity field is expressed as the response of the linearized Navier--Stokes operator to nonlinear forcing,
\begin{equation}
\hat{u} = \mathcal{R}(\omega)\,\hat{f},
\end{equation}
where $\mathcal{R}(\omega)$ is the resolvent operator. The dominant structures are associated with the leading singular modes of $\mathcal{R}$.

\subsubsection*{Identification of the oscillator.}

In the present framework, the internal variables $(A,B)$ represent the reduced dynamics of the dominant singular mode. The pair of complex-conjugate poles identified in the frequency response corresponds to the local oscillator with frequency $\omega(y)$ and damping rate $\sigma(y)$. Thus, the oscillator defined at the mean-field level can be interpreted as the projection of the resolvent dynamics onto its leading mode.

\subsubsection*{Resolved and unresolved components.}

The decomposition into $(A,B,C)$ naturally separates:
\begin{itemize}
\item the resolved coherent dynamics, captured by the dominant mode $(A,B)$,
\item the unresolved background cascade, encoded in the geometric component $C$.
\end{itemize}
This provides a systematic interpretation of the separation between coherent structures and background turbulence.

\subsubsection*{Spatial modulation.}

In the channel geometry, the properties of the resolvent operator depend on $y$ through the mean shear. The present formulation captures this effect through the spatial variation of the oscillator parameters:
\begin{equation}
\omega(y),\quad \sigma(y),\quad |Z(y)|.
\end{equation}
Thus, the dominant mode is not global, but locally modulated by the mean flow.

\subsubsection*{Interpretation.}

The connection established here shows that the mean-field formulation provides a reduced, closed description of the dominant resolvent mode and its interaction with the cascade. In this sense, the present framework can be viewed as a nonlinear, spatially resolved counterpart of resolvent analysis.

\subsubsection*{Remark.}

This connection opens the possibility of systematically linking the present theory with data-driven modal decompositions, and of using resolvent-based insights to refine the structure of the kernel in future developments.

\subsection{Wall-unit estimates of the characteristic length and time scales}

To connect the local oscillator description with wall-resolved numerical simulations, it is useful to express the characteristic spatial and temporal scales in wall units.

\subsubsection*{Wall variables.}

We introduce the standard wall scaling
\begin{equation}
y^+ = \frac{u_\tau y}{\nu},
\qquad
U^+ = \frac{U}{u_\tau},
\qquad
K^+ = \frac{K}{u_\tau^2},
\qquad
\ell^+ = \frac{u_\tau \ell}{\nu},
\qquad
\Omega^+ = \frac{\nu\,\Omega}{u_\tau^2},
\end{equation}
where \(u_\tau=\sqrt{\tau_w}\) is the friction velocity.
The local oscillator scales are then
\begin{equation}
L_{\mathrm{osc}}^+ = \ell^+,
\qquad
T_{\mathrm{osc}}^+ = \frac{1}{\Omega^+}.
\end{equation}

\subsubsection*{Viscous wall layer.}

From the near-wall asymptotic structure,
\begin{equation}
\ell \sim \ell_1 y,
\qquad
K \sim K_2 y^2,
\qquad
\Omega \sim \frac{\sqrt{K_2}}{\ell_1} = \mathrm{const.}
\end{equation}
Therefore,
\begin{equation}
\ell^+ \sim \ell_1\,y^+,
\qquad
K^+ \sim K_2 (y^+)^2,
\qquad
\Omega^+ \sim \Omega_w^+ = \mathrm{const.}
\end{equation}
and hence
\begin{equation}
L_{\mathrm{osc}}^+ \sim y^+,
\qquad
T_{\mathrm{osc}}^+ \sim \mathrm{const.}
\end{equation}
Thus, in the viscous wall layer the oscillator length collapses linearly with wall distance, while its characteristic time remains finite at leading order.

\subsubsection*{Buffer layer.}

In the buffer region, the local forcing and the internal cascade dynamics become comparable,
\begin{equation}
\Pi_S = \frac{S}{\Omega} = O(1),
\qquad
\Pi_B = O(1).
\end{equation}
This implies that the local shear time and the oscillator time are of the same order,
\begin{equation}
T_S \sim T_{\mathrm{osc}}.
\end{equation}
Since the local length scale is still controlled primarily by the wall distance,
\begin{equation}
\ell^+ = O(y^+),
\end{equation}
the buffer region is characterized by
\begin{equation}
L_{\mathrm{osc}}^+ = O(y^+),
\qquad
T_{\mathrm{osc}}^+ = O(y^+),
\end{equation}
up to order-one prefactors set by the local oscillator state.

This provides a natural interpretation of the buffer layer as the region in which the characteristic time scale of the coherent cycle grows with wall distance while remaining strongly coupled to the mean shear.

\subsubsection*{Logarithmic region.}

In the logarithmic layer, the local shear--cascade balance reaches its universal asymptotic form,
\begin{equation}
\frac{S}{\Omega} \to \kappa^{-1},
\qquad
\frac{\mathcal P}{\varepsilon} \to 1.
\end{equation}
Since
\begin{equation}
S \sim \frac{u_\tau}{\kappa y},
\end{equation}
it follows that
\begin{equation}
\Omega \sim \frac{u_\tau}{y},
\end{equation}
or in wall units,
\begin{equation}
\Omega^+ \sim \frac{1}{y^+}.
\end{equation}
Therefore,
\begin{equation}
T_{\mathrm{osc}}^+ \sim y^+,
\qquad
L_{\mathrm{osc}}^+ \sim y^+.
\end{equation}
Thus, in the logarithmic region both the characteristic time and the characteristic length of the oscillator grow linearly in wall units.

\subsubsection*{Summary of regional scaling.}

The three wall regions can therefore be summarized as follows:
\begin{align}
&\text{viscous layer:}
&&L_{\mathrm{osc}}^+ \sim y^+,
\qquad
T_{\mathrm{osc}}^+ \sim \mathrm{const.},
\\[0.3em]
&\text{buffer layer:}
&&L_{\mathrm{osc}}^+ \sim y^+,
\qquad
T_{\mathrm{osc}}^+ \sim y^+,
\\[0.3em]
&\text{logarithmic layer:}
&&L_{\mathrm{osc}}^+ \sim y^+,
\qquad
T_{\mathrm{osc}}^+ \sim y^+.
\end{align}

The essential distinction between the buffer and logarithmic regions is therefore not the scaling itself, but the dynamical state of the oscillator: in the buffer layer the system remains strongly non-equilibrium and maximally active, while in the logarithmic layer it approaches the universal shear--cascade balance.

\subsubsection*{Connection with DNS.}

These estimates provide direct observables for comparison with wall-resolved numerical simulations. In particular, the model predicts:
\begin{itemize}
\item a linear growth of the characteristic oscillator length in wall units,
\item a finite oscillator time at the wall,
\item a crossover toward \(T_{\mathrm{osc}}^+ \sim y^+\) outside the viscous layer,
\item maximal internal activity in the buffer region, where \(\Pi_S\) and \(\Pi_B\) are both order unity.
\end{itemize}
This makes it possible to compare the local state of the oscillator with the time scales and coherence lengths associated with near-wall bursting events and elongated wall structures in DNS.

\subsubsection*{Local time scales and the self-sustaining process.}

The oscillator structure implies the existence of three distinct local time scales:
\begin{equation}
T_{\mathrm{osc}}(y) = \frac{2\pi}{\omega(y)},
\qquad
T_{\mathrm{rel}}(y) = \frac{1}{\sigma(y)},
\qquad
T_{\mathrm{burst}}(y) \sim \frac{1}{S(y)}.
\end{equation}

These correspond, respectively, to the internal oscillation of the cascade, the relaxation of the dominant mode, and the characteristic time of shear-driven energy injection.

\subsubsection*{Connection with the self-sustaining process.}

These three time scales provide a minimal dynamical representation of the near-wall self-sustaining process (SSP):
\begin{itemize}
\item shear-driven amplification (time scale $1/S$),
\item internal oscillation of coherent structures (time scale $T_{\mathrm{osc}}$),
\item relaxation and breakdown (time scale $T_{\mathrm{rel}}$).
\end{itemize}

\subsubsection*{Wall-layer interpretation.}

The relative ordering of these time scales determines the dynamical regime:
\begin{itemize}
\item near the wall, $S/\Omega \ll 1$, the oscillator is weakly forced and no full cycle develops,
\item in the buffer region, $S/\Omega = O(1)$, the three time scales become comparable and the SSP is fully active,
\item in the logarithmic region, $S/\Omega \to \kappa^{-1}$, the cycle reaches a statistically stationary regime.
\end{itemize}

\subsubsection*{Interpretation.}

This shows that the self-sustaining process is not an additional ingredient, but is already encoded in the local oscillator structure of the mean-field formulation. The channel flow organizes this dynamics spatially, selecting the regime of the oscillator at each wall-normal location.

\end{document}